\documentclass[]{emulateapj}

\usepackage{comment}
\usepackage{subfigure}
\usepackage{url}

\shorttitle{Chemical Abundances in Pal 1}
\shortauthors{Sakari et al.}

\begin{document}

\title{Detailed Chemical Abundances of Four Stars in the Unusual Globular
  Cluster, Palomar 1}
\author{Charli M. Sakari, Kim A. Venn}
\affil{Department of Physics and Astronomy, University of Victoria,
  Victoria, BC V8W 3P2, Canada}
\email{sakaricm@uvic.ca, kvenn@uvic.ca}

\author{Mike Irwin}
\affil{Institute of Astronomy, University of Cambridge, Madingley
  Road, Cambridge CB3 0HA, UK}
\email{mike@ast.cam.ac.uk}

\author{Wako Aoki\altaffilmark{1}, Nobuo Arimoto\altaffilmark{1}}
\affil{National Astronomical Observatory of Japan, Mitaka, Tokyo
  181-8588, Japan}
\email{aoki.wako@nao.ac.jp, arimoto.n@nao.ac.jp}

\author{Aaron Dotter}
\affil{Space Telescope Science Institute, Baltimore, MD 21218, USA}
\email{dotter@stsci.edu}

\altaffiltext{1}{Department of Astronomical Science, The Graduate
University for Advanced Studies, Mitaka, Tokyo 181-8588, Japan}

\begin{abstract}
Detailed chemical abundances for twenty one elements are presented for
four red giants in the anomalous outer halo globular cluster Palomar 1
($R_{\rm{GC}} = 17.2$ kpc, $Z=3.6$ kpc) using high-resolution
($R=36000$) spectra from the High Dispersion Spectrograph (HDS) on the
Subaru Telescope. Pal 1 has long been considered unusual because of
its low surface brightness, sparse red giant branch, young age, and
its possible association with two extragalactic streams of
stars---this paper shows that its chemistry further confirms its
unusual nature.  The mean metallicity of the four stars, $[\rm{Fe/H}]
= -0.60 \pm 0.01$, is high for a globular cluster so far from the
Galactic center, but is low for a typical open cluster.  The
[$\alpha$/Fe] ratios, though in agreement with the Galactic stars
within the $1\sigma$ errors, agree best with the lower values in dwarf
galaxies.  No signs of the Na/O anticorrelation are detected in Pal 1,
though Na appears to be marginally high in all four stars.  Pal 1's
neutron capture elements are also unusual: its high [Ba/Y] ratio
agrees best with dwarf galaxies, implying an excess of second-peak
over first-peak s-process elements, while its [Eu/$\alpha$] and
[Ba/Eu] ratios show that Pal 1's contributions from the r-process must
have differed in some way from normal Galactic stars.   Therefore, Pal
1 is chemically unusual, as well in its other properties.  Pal 1
shares some of its unusual abundance characteristics with the young
clusters associated with the Sagittarius dwarf galaxy remnant and the
intermediate-age LMC clusters, and could be chemically associated with
the Canis Majoris overdensity; however it does not seem to be similar
to the Monoceros/Galactic Anticenter Stellar Stream.
\end{abstract}

\keywords{galaxies: dwarf --- globular clusters: general ---
globular clusters: individual(Pal 1)}

\section{Introduction}\label{sec:Intro}
\begin{figure}
\epsscale{1.2}
\plotone{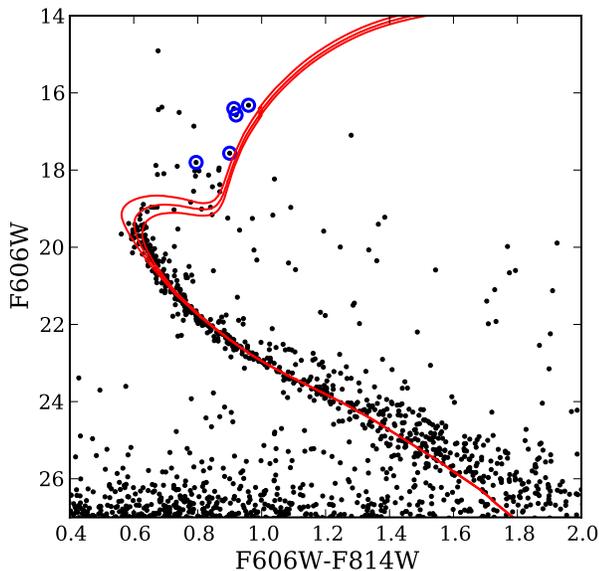}
\caption{An HST color-magnitude diagram (roughly V, V-I) for the
central field of Pal 1 from \citet{Sarajedini}.  The observed targets
are circled.  Also shown are fitted isochrones from the Dartmouth
Stellar Evolution Database \citep{DSED} for ages of 4, 5, and 6 Gyr.
Values of $[\rm{Fe/H}] = -0.6$, $[\alpha/\rm{Fe}]=0.0$, $(m-M)_{V} =
16.27$, and $E(B-V) = 0.20$ are adopted, as discussed in the
text\label{fig:CMD}}
\end{figure}

Ever since its discovery by \citet{Abell}, Palomar 1 (Pal 1) has been
tentatively classified as a globular cluster (GC), primarily because
of its location high above the Galactic plane ($Z=3.6$ kpc;
\citealt{Harris}, 2010 edition).  However, Pal 1 has several anomalous
characteristics that have cast doubt upon this classification. For
example, Pal 1 appears to be unusually young for an outer halo GC.
Using standard isochrone fits to Pal 1's ground-based color-magnitude
diagram (CMD), \citet{RosenbergA} estimated an age between $6.3$ and 
$8$ Gyr, while \citet{Sarajedini} found an age between 4 and 6 Gyr
with HST photometry.  This F606W ($\sim$V), F814W ($\sim$I) HST CMD of
the central field of Pal 1 from \citealt{Sarajedini} (see Figure
\ref{fig:CMD}) shows the main sequence turn-off around F606W $\approx
19.4$ ($V\approx 19.6$), and a possible detection of a red horizontal
branch near F606W $\approx 16.4$ ($V\approx 16.6$).  Thus, $\Delta
V^{\rm{HB}}_{\rm{MSTO}} \approx 3$, making the cluster younger than
about 7 Gyr (with the method of \citealt{Chaboyer}).  These estimates
make Pal 1 one of the youngest GCs, along with Ter 7, Pal 12, and
Whiting 1, whose ages range from $7-9$ Gyr
(e.g. \citealt{Buonanno1998}; \citealt{SalarisWeiss};
\citealt{Carraro}; \citealt{Dotter2010}).

Pal 1 also lacks luminous, evolved stars.  The CMD in Figure
\ref{fig:CMD} shows a sparsely populated red giant branch and a barely
detectable horizontal branch.  Furthermore, near-infrared
medium-resolution spectroscopy of the \ion{Ca}{2} triplet (CaT) lines
in four of these red giants has shown that Pal 1 is fairly metal-rich
for an outer halo GC, with an average $[\rm{Fe/H}] = -0.6$ (on the
Zinn-West scale, using the standard Galactic CaT line strength to
metallicity conversion; \citealt{RosenbergB}).  The main sequence is
also poorly populated, and therefore the entire cluster has an
extremely low total luminosity for a GC ($M_V = -2.52$;
\citealt{Harris}, 2010 edition).

Clearly Pal 1 does not conform to the characteristics of traditional
Galactic GCs, suggesting either that Pal 1 is not a GC or that it is
not Galactic.  If Pal 1 has been misclassified as a GC, another
possibility is that it could be an open cluster (OC)---this option
seems unlikely due to the combination of its position above the
Galactic plane, its high concentration parameter ($c=2.57$;
\citealt{Harris}, 2010 edition), and its age (although Pal 1's age
\textit{is} comparable to that of the oldest known OCs,
e.g. \citealt{VandenbergStetson}).  However, Pal 1's brightness and
age place it in greater agreement with the OCs than with the GCs (see
Figure 3 in \citealt{Carretta2010}).  The ambiguity surrounding the OC
vs. GC classification leads to a need for a more rigorous distinction
between the two.  From observations of many Galactic GCs,
\citet{Carretta2010} defined a GC as a cluster that shows signs of a
second generation of stars through a Na/O anticorrelation.  Under this
definition, if Pal 1 does \textit{not} show signs of a Na/O
anticorrelation, then it cannot be considered a classical
GC.\footnote{In the past, Pal 1's unusual characteristics have led to
a designation as a \textit{transitional} cluster, i.e. an object that
resembles something between classical globular and open clusters,
e.g., \citet{Ortolani}.}

Alternatively, Pal 1 could have originated in a
dwarf galaxy and been accreted by the Milky Way during a merger.
The latter situation is plausible, as several other Galactic GCs,
notably the young clusters listed above, have been linked to dwarf
galaxies.  M54, Arp 2, Ter 7, and Ter 8 lie on a stream associated
with the Sagittarius dwarf spheroidal (Sgr dSph) galaxy
\citep{DaCostaArmandroff}; Pal 12 has been kinematically linked to the
Sgr dSph through proper motion studies \citep{Dinescu}; Whiting 1 is
undergoing tidal stripping and appears to lie in a stream of M giants
from the Sgr dSph \citep{Carraro}; and Rup 106 has been tentatively
associated with the Magellanic Clouds because of its location along
the Magellanic stream \citep{LinRicher}.  Pal 1 does appear to lie
along several stellar streams that may be extragalactic but have not
yet been associated with any galaxies.  One such stream, known both as
the Monoceros stream and as the Galactic Anticenter Stellar Stream
(GASS), appears to be from a disrupted dwarf satellite
(e.g. \citealt{Sollima}); Pal 1's position, velocity, and metallicity
make it a possible member of this stream \citep{Crane}.  The Canis
Major overdensity, which seems to be distinct from the GASS
\citep{Chou}, has also been associated with Pal 1, again because of
the cluster's coincidental location \citep{Martin}. Finally, Pal 1 has
also been tentatively linked to the ``Orphan Stream''
\citep{Belokurov}, though its radial velocity and metallicity make it
an unlikely member \citep{Newberg}. Though no obvious association with
a particular stream has been confirmed, tidal tails around Pal 1 have
recently been discovered, suggesting that the cluster itself is being
dissipated in the tidal field of the Galaxy \citep{NiedersteOstholt}.

In the absence of an obvious host galaxy, detailed chemical abundances
can be used to examine if Pal 1 has an extragalactic origin.  It has
been well established (e.g. \citealt{Tolstoy}, \citealt{Venn2004})
that metal-rich stars in Local Group dwarf galaxies have distinctly
different chemical abundances from stars in the Milky Way, at a given
metallicity (the differences are not so obvious by $[\rm{Fe/H}] \sim
-2$).  These chemical variations between the environments are due to
different star formation histories and efficiencies, which are linked
to the mass of the galaxy and the properties of its environment.
Therefore, the differences are not just between dwarf galaxies and
normal galaxies---in fact, dwarf galaxies have unique chemical
evolutions, meaning that it is possible to chemically link a star to
its host galaxy (e.g. \citealt{Tolstoy}, \citealt{Freeman}). These
chemical differences are not just seen in field stars, they are also
seen in GCs; for example, \citet{Pritzl} showed that most Galactic GCs
have similar chemical abundances to Galactic field stars at the same
metallicity, with Pal 12, Ter 7, and Rup 106 as notable exceptions.

Detailed, high-resolution chemical abundances have been used to trace
the origin of several globular clusters.  In particular Pal 12
\citep{Cohen} and Ter 7 (\citealt{Tautvaisiene}, \citealt{Sbordone})
have [$\alpha$/Fe] ratios that follow the Sgr dSph field star
abundance trends and are clearly separated from Galactic field stars,
supporting the hypothesis that they were captured from the Sgr dSph.
The more metal-poor Sgr clusters Arp 2 and Ter 8, however, are not
chemically distinct from the Galaxy \citep{Mottini}, which shows that
the [$\alpha$/Fe] deviation can be metallicity- (and presumably age-)
dependent.  If Pal 1 has an extragalactic origin, then its metallicity
of $[\rm{Fe/H}]\approx -0.6$ puts it in a regime where [$\alpha$/Fe]
and other chemical abundance ratios should be useful in determining
its origin.

Here we present a detailed analysis of high-resolution spectra for
four red giant stars in Pal 1, taken with the Subaru Telescope.
\citet{MonacoPal1} have completed a chemical analysis of one star in
Pal 1 (Pal 1-I), also with spectra taken at Subaru, but with
independent observations; they find that their single star shows
similar abundance patterns to Galactic open clusters. \citet{Saviane}
further show that its chemistry implies that Pal 1 could be associated
with the Canis Major overdensity.  With four stars, our analysis
allows us to determine the chemical abundance ratios with higher
precision, and investigate any star-to-star variations in the
abundances that could be linked with chemical or cluster evolution 
effects. Section \ref{sec:Observations} outlines the observations and
the data reduction, while Section \ref{sec:EWs} examines our method
for measuring equivalent widths.  In Section \ref{sec:ModelAtms} we
discuss the model atmospheres, particularly how the atmospheric
parameters are derived. Section \ref{sec:Abundances} presents the
methods for element abundance determinations, while Section
\ref{sec:Results} examines the significance of these results with
respect to Galactic stars and clusters. Finally, Section
\ref{sec:Discussion} examines the origins of Pal 1.

\section{Observations and Data Reduction}\label{sec:Observations}
\begin{figure}
\epsscale{1.0}
\plotone{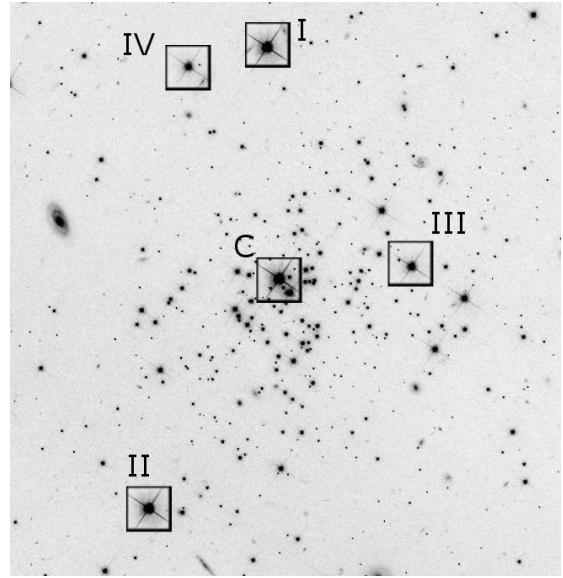}
\caption{An HST F814W image of Pal 1, with our targets shown in
squares.  The field of view is approximately
$1.\arcmin7 \times 1.\arcmin7$.  North is up, and west is to the left.
Note that we do not perform a full analysis of Pal 1-IV due to poor
sky conditions.\\ \label{fig:targets}}
\end{figure}

Four of our target stars are the same stars observed by
\citet{RosenbergB}: bright, red giant branch (RGB) stars, with radial
velocities that make them likely cluster members.  We refer to these
stars as Pal 1-I, -II, -III, and -IV, following the naming convention
of Rosenberg et al.  The bright object located at the center of the
cluster, Pal 1-C, was also included as a target, although the
membership of this object is not as certain as the others.  The
locations of these stars on a CMD and in an image of the cluster (from
\citealt{Sarajedini}) are shown in Figures \ref{fig:CMD} and
\ref{fig:targets}.  In addition to the Pal 1 stars we include an
analysis of a ``standard'' star, M67-141 (as identified by
\citealt{Fagerholm}).  This star resides in the metal-rich open
cluster M67, and has previous detailed, high resolution chemical
abundances from \citet{Yong} and \citet{Pancino}.  Table
\ref{table:targets} shows the positions and magnitudes of the target
stars.  The Pal 1 positions and V and I magnitudes (in the Johnson
system) are from \citet{Sarajedini}, while M67-141's position and V
and I magnitudes (in the Kron-Cousins system) are from \citet{Hog},
\citet{Sanders}, and \citet{JanesSmith}, respectively. All of the K
magnitudes are from the Two Micron All-Sky Survey (2MASS) Point Source
Catalog.\footnote{\url{http://tdc-www.harvard.edu/catalogs/tmpsc.html}}

\begin{deluxetable*}{lcccccc}
\tabletypesize{\scriptsize}
\tablecolumns{7}
\tablewidth{0pc}
\tablecaption{Our target stars, their positions, and their magnitudes\label{table:targets}}
\tablehead{
Star & RA (J2000) & Dec (J2000) & V\tablenotemark{a} & I\tablenotemark{a} & K\tablenotemark{b} & References\tablenotemark{c}}
\startdata
M67-141 & $8^{\rm{h}}51^{\rm{m}}22.8^{\rm{s}}$ & $+11^{\circ}48\arcmin1.\arcsec7$ & 10.480 & 9.400 & 7.942 & 1, 2, 3\\
Pal 1-I & $3^{\rm{h}}33^{\rm{m}}21.8^{\rm{s}}$ & $+79^{\circ}35\arcmin16.\arcsec2$ & 16.675 & 15.459 & 13.832 & 4\\
Pal 1-II & $3^{\rm{h}}33^{\rm{m}}29.6^{\rm{s}}$ & $+79^{\circ}34\arcmin16.\arcsec2$ & 16.843 & 15.618 & 13.983 & 4\\
Pal 1-III & $3^{\rm{h}}33^{\rm{m}}12.3^{\rm{s}}$ & $+79^{\circ}34\arcmin59.\arcsec2$ & 17.827 & 16.628 & 15.281 & 4\\
Pal 1-IV & $3^{\rm{h}}33^{\rm{m}}27.0^{\rm{s}}$ & $+79^{\circ}35\arcmin34.\arcsec9$ & 18.032 & 16.969 & 15.779 & 4\\
Pal 1-C & $3^{\rm{h}}33^{\rm{m}}21.0^{\rm{s}}$ & $+79^{\circ}34\arcmin57.\arcsec1$ & 16.603 & 15.328 & 13.715 & 4\\
\enddata
\tablenotetext{a}{M67-141 V and I magnitudes are in the Kron-Cousins
system, while the Pal 1 V and I magnitudes have been transformed to
the Johnson system from the HST system.}
\tablenotetext{b}{All K magnitudes are from the 2MASS Point Source
Catalog.}
\tablenotetext{c}{\textbf{References.} (1) \citet{Hog}; (2)
\citet{Sanders}; (3) \citet{JanesSmith}; (4) \citet{Sarajedini}}
\end{deluxetable*}

These stars were observed on several runs in 2006 January and 2007
January using the High Dispersion Spectrograph (HDS; \citealt{HDSref})
on the 8.2 m Subaru Telescope.  The default grating was centered at
5500 \AA \hspace{0.025in} with a slit length of 5.6 arcsec for all
stars and a slit width of 1 arcsec for the Pal 1 stars and 0.6 arcsec
for M67-141, leading to spectral resolutions of $R=36000$ and
$R=60000$, respectively.  Table \ref{table:observations} shows a list
of the dates, total exposure times, and signal-to-noise ratios (SNR)
for these observations.  We note that the position of the Moon was not
preferable for observations of the Pal 1 stars, as the primary target
of this observing run was the Sextans dwarf galaxy \citep{Aoki2009}.

The data were reduced using the basic Subaru HDS pipeline, with
standard IRAF\footnote{IRAF is distributed by the National Optical
Astronomy Observatory, which is operated by the Association of
Universities for Research in Astronomy, Inc., under cooperative 
agreement with the National Science Foundation.} routines to remove
the bias and flat field, and with additional cosmic ray removal.  The
stars were fairly low in the sky during the observations and the
seeing conditions were poor, meaning the stars observed in 2006 nearly
filled the slit; as a result the sky subtractions proved to be
extremely difficult.  Ultimately some of the star signals were removed
along with the sky even when using the outermost 3-pixels on either
side of the projected slit spectrum; thus, the final, co-added stellar
spectra from 2006 (stars Pal 1-I and -II) have lower SNR than
expected.  The sky subtraction for one of the 2007 stars, Pal 1-IV,
proved impossible, and that star has therefore been dropped from
further abundance analyses.  The other two stars, Pal 1-III and -C,
have good sky subtraction and therefore higher SNR.

The radial velocities were determined by locating several strong,
easily-identified spectral lines and calculating the shift in
wavelength using Gaussian fits to the lines. These radial velocities
were then shifted to heliocentric velocities using the IRAF task
\textit{rvcorrect}; the final heliocentric velocities are listed in
Table \ref{table:observations}.  Though a sky subtraction could not be
performed on Pal 1-IV, a radial velocity could still be obtained; this
radial velocity is quite different from the other stars, as will be
discussed in Section \ref{subsubsec:UFD}.

\begin{center}
\begin{deluxetable*}{llcccc}
\tabletypesize{\scriptsize}
\tablecolumns{6}
\tablewidth{0pc}
\tablecaption{Dates of observations, total exposure times,
signal-to-noise ratios at two wavelengths, and calculated heliocentric
radial velocities \label{table:observations}}
\tablehead{
Star & Dates & Exposure Time (s) & SNR\tablenotemark{a} (5310 \AA) & SNR\tablenotemark{a} (6616
\AA) & $v_{\rm{helio}}$ (km s$^{-1}$)}
\startdata
M67-141 & 2006 Jan 21 & 5985     & 100 & 150 & $+33.9\pm 1.0$ \\
Pal 1-I & 2006 Jan 20, 21 & 9000 & 15 & 30 & $-77.2\pm 1.0$ \\
Pal 1-II & 2006 Jan 20 & 9000    & 15 & 30 & $-78.0\pm 1.0$ \\
Pal 1-III & 2007 Jan 25, 26 & 12600 & 31 & 45 & $-77.2\pm 1.0$ \\
Pal 1-IV & 2007 Jan 27, 28, 29 & 21600 & -\tablenotemark{b} & -\tablenotemark{b} & $-68.1\pm 1.0$ \\
Pal 1-C & 2007 Jan 6, 27 & 5400 & 35 & 55 & $-77.0\pm 1.0$ \\
\enddata
\tablenotetext{a}{SNR are per resolution element.}
\tablenotetext{b}{Pal 1-IV was dropped from the analysis, as the
seeing was too poor to perform a sky subtraction.\\}
\end{deluxetable*}
\end{center}

\begin{figure}
\begin{center}
\subfigure[Blue Region]{\includegraphics[scale=0.5]{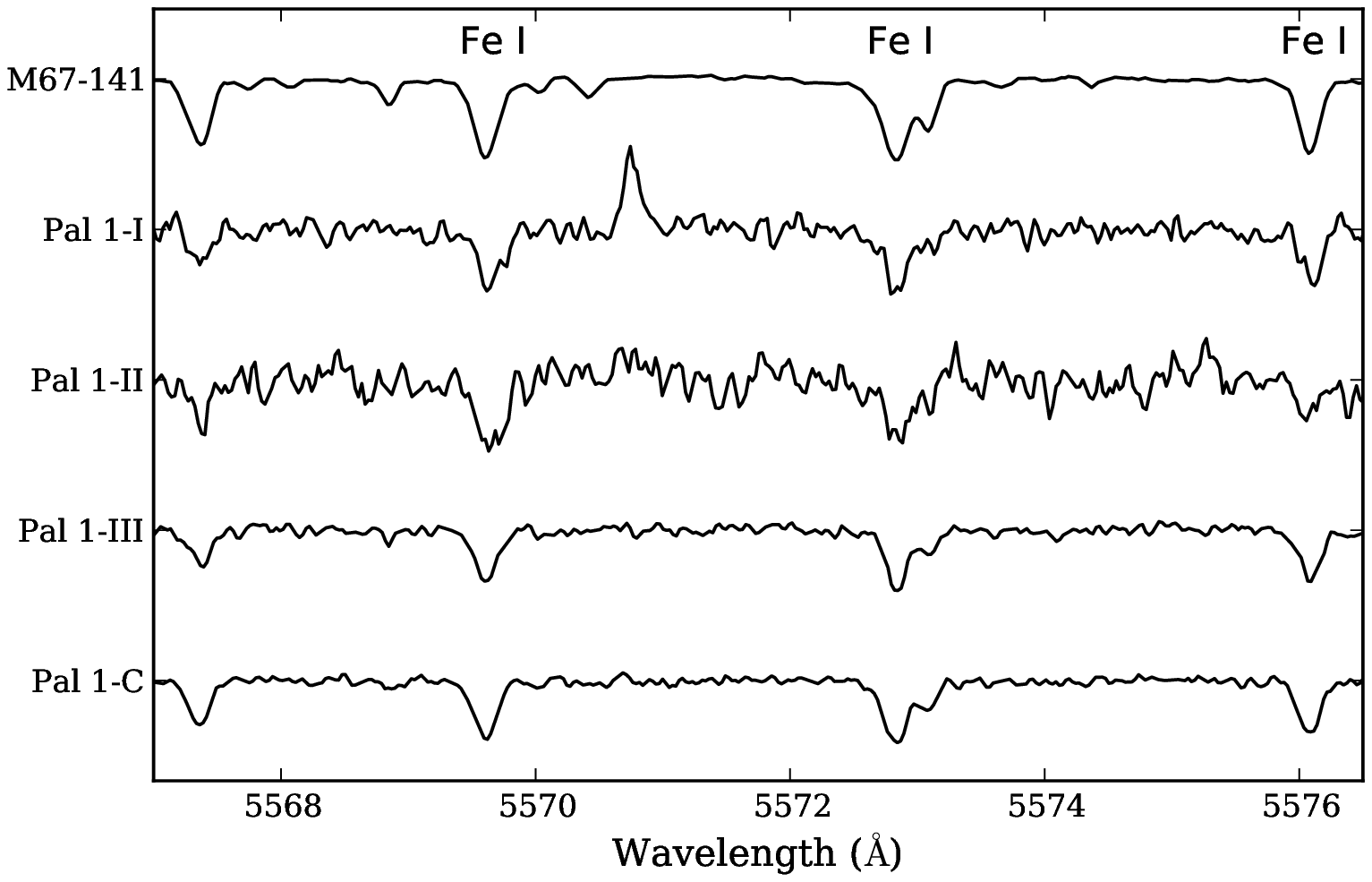}}
\subfigure[Red Region]{\includegraphics[scale=0.5]{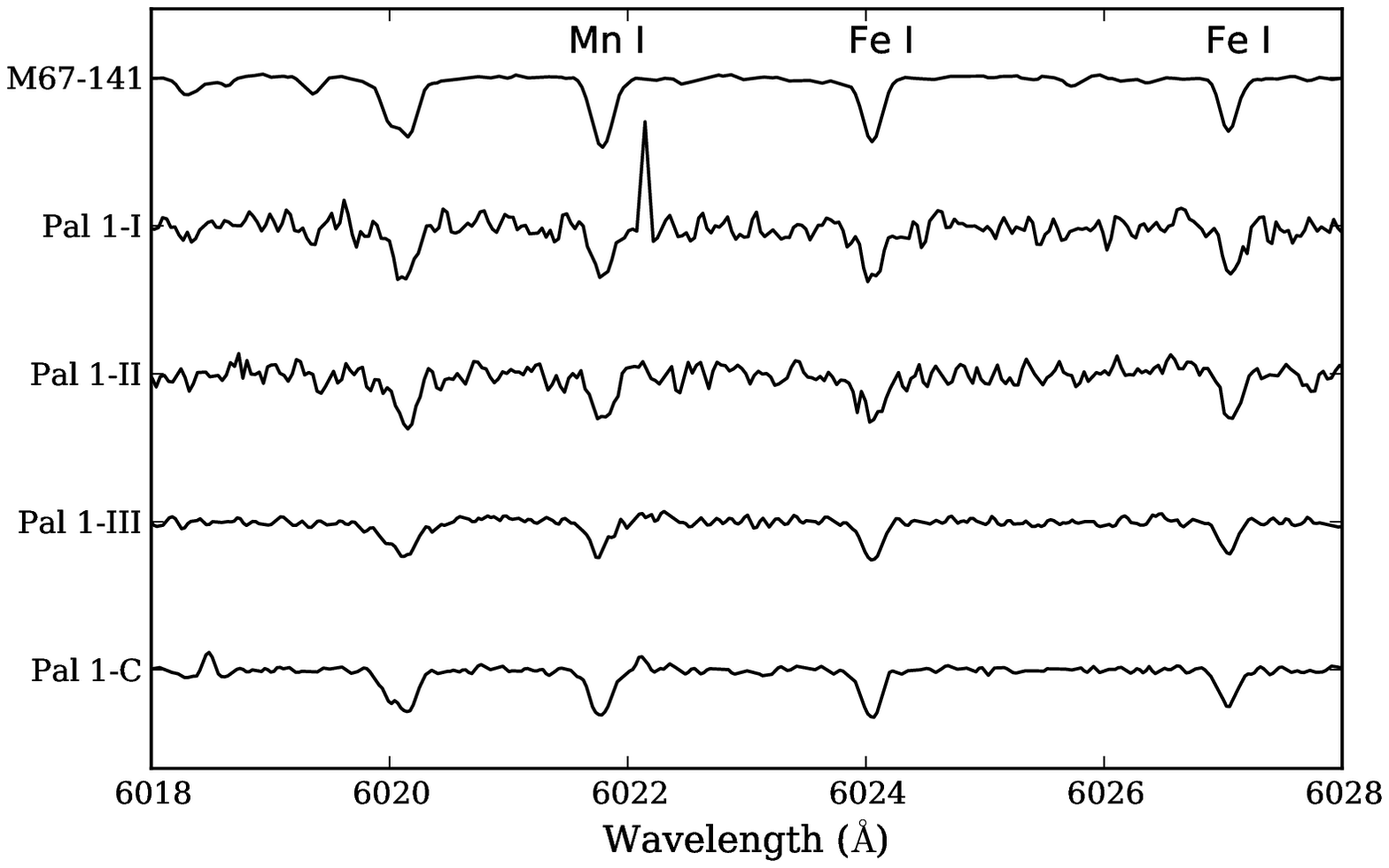}}
\caption{Portions of the five spectra, in both a blue and a red
region, with notable spectral lines labelled.  The spectra have been
continuum normalized and were arbitrarily shifted vertically for the
ease of comparison.  Note the low S/N of Pal 1-I and -II compared to
the other Pal 1 stars.\label{fig:spectra}}
\end{center}
\end{figure}

\vspace{-0.4in}
The shifted rest-frame spectra were then continuum-normalized.  The
continuum was estimated using an iterative, non-parametric filter
(e.g. \citealt{Battaglia}) designed to split the spectrum into
continuum and line components based on a user-defined scale length for
allowed continuum variations.  This was set at 10 \AA \hspace{0.025in}
so that the strongest spectral absorption lines were not affected.
Figure \ref{fig:spectra} shows red and blue portions of the final
stellar spectra which have a full coverage of approximately 4350 to
7140 \AA; the representative spectra in Figure \ref{fig:spectra} have
been arbitrarily shifted vertically for easier comparison.

\section{Equivalent Widths}\label{sec:EWs}
The measured spectral lines are from the line list by
\citet{Shetrone}, with supplemental lines from \citet{Cayrel2004},
\citet{Aoki2007}, \citet{Cohen2008}, \citet{Letarte2009},
\citet{Tafelmeyer2010}, and \citet{Frebel2010}.  The equivalent widths
(EWs) were fit with Gaussian profiles using \textit{splot} in IRAF and
were checked by numerical interpretation under the continuum.   EWs
can now be measured with automatic line-measuring programs such as
DAOSPEC,\footnote{DAOSPEC has been written by P.B. Stetson for the
Dominion Astrophysical Observatory of the Herzberg Institute of
Astrophysics, National Research Council, Canada.} a program that is
intended for high resolution ($R>15000$), high SNR ($>30$) spectra
\citep{DAOSPECref}.  Some advantages of DAOSPEC over IRAF's
\textit{splot} include a fixed full-width at half maximum (FWHM) for
all lines, and an effective continuum that takes weak features into
account; however, it appears that with low SNR spectra DAOSPEC can
have difficulty fitting the continuum.  Due to the poor quality of the
Pal 1-I and -II spectra, IRAF \textit{splot} measurements are
therefore preferred.  By a visual comparison with the higher SNR
spectra (Pal 1-III and -C), the spectral lines in Pal 1-I and -II were
identified and distinguished from noise in the low SNR
spectra. Spectral lines in the noisy spectra that had a drastic
difference in width or depth from Pal 1-III or -C were discarded.
\begin{center}
\begin{deluxetable*}{lccccccccc}
\centering
\tabletypesize{\scriptsize}
\tablecolumns{10}
\tablewidth{0pc}
\tablecaption{The line list.\tablenotemark{a}\label{table:LineList}}
\tablehead{
Wavelength & Element & E.P. & log gf & \multicolumn{6}{l}{Equivalent width (m\AA)}\\
(\AA) & & (eV) & & Sun & M67-141 & Pal 1-I & Pal 1-II & Pal 1-III & Pal 1-C}
\startdata
 4443.19 &  \ion{Fe}{1} &  2.86 & -1.043 &  134.0 & 171.0 & - & - & - & -\\
 4476.02 &  \ion{Fe}{1} &  2.85 & -0.819 &  - & - & - & - & - & 157.0\\
 4484.22 &  \ion{Fe}{1} &  3.60 & -0.864 &  100.0 & 122.0 & - & - & 76.0 & 117.0\\
 4489.75 &  \ion{Fe}{1} &  0.12 & -3.899 &  91.0 & - & - & - & 120.0 & -\\
 4592.66 &  \ion{Fe}{1} &  1.56 & -2.462 &  99.0 & 185.0 & - & - & - & -\\
\enddata
\tablenotetext{a}{Table \ref{table:LineList} is published in its
entirety in the electronic edition of \textit{The Astrophysical
Journal}. A portion is shown here for guidance regarding its form and
content.\\}
\end{deluxetable*}
\end{center}

\vspace{-0.35in}
Measurements were made relative to the local continuum.  For all
stars, lines stronger than 200 m\AA \hspace{0.025in} were rejected.
For Fe, the Pal 1 lines with EWs stronger than 150 m\AA
\hspace{0.025in} were also thrown out, in order to best constrain the
atmospheric parameters (discussed in Section \ref{sec:ModelAtms}).
The wavelengths (in \AA), excitation potentials (in eV), $\log gf$
values, and equivalent widths (in m\AA) for the four Pal 1 stars and
the standard star are shown in Table \ref{table:LineList}.

\subsection{IRAF \textit{splot} vs. DAOSPEC measurements}\label{subsec:HandvsDAOSPEC}
\begin{figure}
\epsscale{1.25}
\plotone{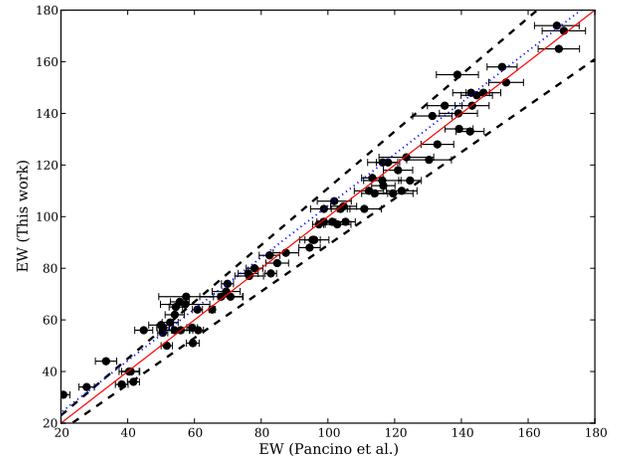}
\caption{A comparison of the EWs as measured with \textit{splot} in
IRAF to those measured by the program DAOSPEC (from \citealt{Pancino})
for the standard star M67-141.  The red line shows equal EW
measurements.  The blue dotted line shows the average offset between
the weak lines in the two data sets.  The dashed lines show the EW
errors ($\rm{EW}_{\rm{min}} + 10\% \; \rm{EW}$) as calculated using
the \citet{Cayrel} formula (see Section
\ref{subsec:errors}).\\ \label{fig:HandvsDAOSPEC}}
\end{figure}

DAOSPEC has been used for spectral line measurements by several
different authors (e.g. \citealt{Letarte2010}); therefore we compare
our \textit{splot} EWs to those from DAOSPEC.  Figure
\ref{fig:HandvsDAOSPEC} shows a comparison between \textit{splot} EWs
(from this work) and DAOSPEC EWs (from \citealt{Pancino}) for the
standard star, M67-141, which has the highest resolution and SNR.
Their EW errors from DAOSPEC are also shown.  The majority of the
offsets between the \textit{splot} and the DAOSPEC measurements are
within our adopted EW measurement errors (shown as dashed lines; see
Section \ref{subsec:errors} for a description of how these errors are
calculated), and are therefore not significant.  The few lines outside
these errors are weak lines that we measure to be slightly stronger
than DAOSPEC measures---this difference may be attributed to continuum
placement (see \citealt{DAOSPECref} for a discussion of their
effective continuum).  For the weak lines ($>60$ m\AA), the average
discrepancy between the two analyses (shown as a dotted blue line) is
$4 \pm 6$ m\AA.

\subsection{A Comparison with Previous Studies}\label{subsec:MonacoEWs}
\begin{figure}
\epsscale{1.25}
\plotone{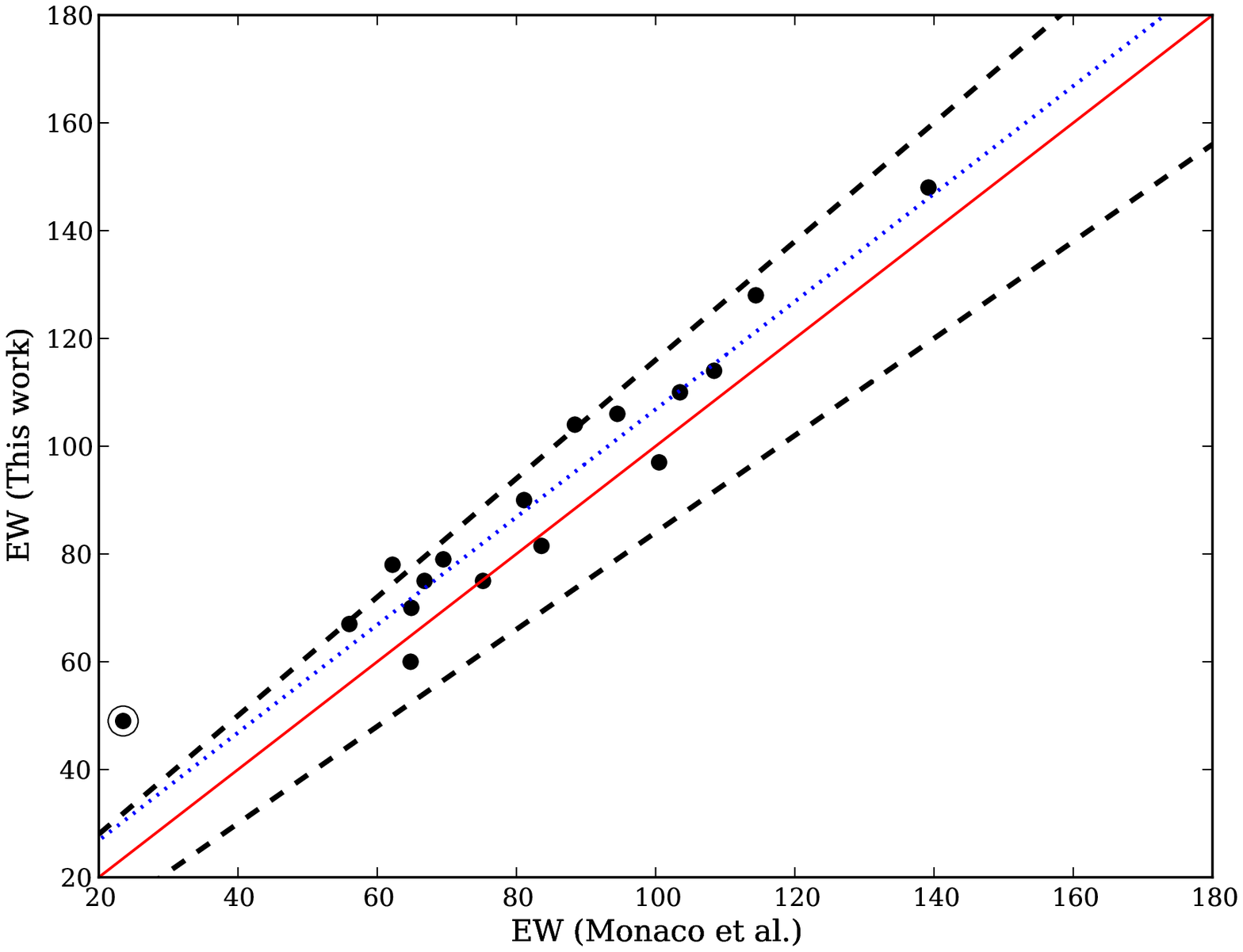}
\caption{A comparison between our EWs for Pal 1-I, and those measured
by \citet{MonacoPal1}, for the lines in common between our analyses.
EW errors were not given in \citet{MonacoPal1}.  The circled point is
a \ion{Cr}{1} line that is unusally discrepant; for that reason we
have removed it from the analysis. The dashed lines represent the EW
errors, as discussed in Section \ref{subsec:errors}. The blue dotted
line shows the average offset between the two data sets.  Note that
though we tend to measure slightly larger EWs, the differences are
typically less than the EW errors.\\ \label{fig:UsVsThem}}
\end{figure}

Our EW measurements from the low SNR Pal 1-I spectrum can be compared
to those from \citet{MonacoPal1}. Unfortunately there are very few
lines in common for us to compare, due to differences in the spectral
range and SNR . For the lines in common, Figure \ref{fig:UsVsThem}
shows our \textit{splot} EWs versus the Monaco et al. DAOSPEC EWs from
the higher SNR spectrum.  Also shown are the line of perfect agreement
(solid red line), our estimated 1$\sigma$ EW errors (black dashed
lines; see Section \ref{subsec:errors}), and the average trend (dotted
blue line).  Our EWs tend to be slightly higher, with an average
offset of $7\pm 6$ m\AA.  However, the majority of the points in
Figure \ref{fig:UsVsThem} lie within the dashed lines, which implies
that the EW measurement techniques and spectral reduction methods are
in reasonable agreement.  We further note that there does not seem to
be a significant trend in error with EW.

\vspace{0.4in}

\section{Model Atmospheres}\label{sec:ModelAtms}
Stellar atmospheric parameters (effective temperature, $T_{\rm{eff}}$;
surface gravity in its logarithmic form, $\log g$; microturbulence,
$\xi$, in km s$^{-1}$; and metallicity, [\ion{Fe}{1}/H]\footnote{We
use the standard notation:
$\rm{[X/H]}=(12+\log(N_{\rm{X}}/N_{\rm{H}})) -
(12+\log(N_{\rm{X},\sun}/N_{\rm{H},\sun})) = \log \epsilon(\rm{X}) -
\log \epsilon(\rm{X})_{\sun} $, where $X$ is any element, and $N_X$
and $N_H$ are the column densities of element X and of hydrogen,
respectively.}) are initially derived from broadband photometry and
revised based on spectral indicators as described below.

\subsection{Photometric Parameters}\label{subsec:PhotometricAtm}
Determining the atmospheric parameters from broadband colors requires
knowledge of the metallicity, reddening, distance modulus, and mass of
the stars.  The initial metallicity for M67-141, $[\rm{Fe/H}]=0.01$,
comes from \citet{Yong}, while the initial metallicity for the Pal 1
stars, $[\rm{Fe/H}]=-0.6$, comes from the CaT measurements of
\citet{RosenbergB}.  The surface temperatures are estimated using the
colors of the stars (from the magnitudes shown in Table
\ref{table:targets}).  The $(V-K)$ color to temperature conversion of
\citet[with the \citealt{AlonsoCorrection} correction]{Alonso} is
used---this conversion is based on a large sample of Galactic GCs and
field stars and is calibrated in the TCS photometric system.
Conversions from $(V-K)_{\rm{Johnson}}$ to $(V-K)_{\rm{TCS}}$ are
therefore used \citep{Alonso1998}.

The \citet{Alonso} calibrations are based on the \textit{absolute}
colors of the stars, meaning the cluster reddening must be taken into
account. The adopted reddenings for Pal 1 and M67-141 are $E(B-V) =
0.20 \pm 0.03$ and $E(B-V) = 0.033 \pm 0.005$, from the
\citet{Schlegel} dust extinction maps.  The $E(V-K)$ reddening is
determined with the conversions of \citet{McCall}.  The final
photometric temperatures are shown in Table \ref{table:AtmParams}.

The bolometric correction for each star is calculated using the
formula from \citet{Alonso}, and the absolute bolometric magnitude is
determined using the distance modulus.  For M67-141 we simply adopt
the absolute distance modulus quoted in \citet{Yong}: $(m-M)_0=9.56$.
For Pal 1 there are two different literature values for the distance
modulus: \citet{RosenbergA} give (m-M)$_0 = 15.25 \pm 0.25$ while
\citet{Sarajedini} give (m-M)$_0 = 15.65 \pm 0.10$.  Both are
determined through main sequence fitting of Pal 1 to the lower main
sequence of 47 Tuc, a Galactic globular cluster with a similar
metallicity to Pal 1.\footnote{As discussed in \citet{RosenbergA}, the
presence of the sparse horizontal branch is uncertain, preventing
accurate determinations of the distance modulus using zero age
horizontal branch methods (e.g. \citealt{Vandenberg2000}).}  We adopt
the \citet{Sarajedini} distance modulus, (m-M)$_V =
16.26$,\footnote{We assume here that $A_V = 3.07 E(B-V)$
\citep{McCall}.} which is determined from HST photometry.

Finally, the photometric surface gravity is determined through a
luminosity comparison with the Sun, which requires knowledge of the
mass of our stars.  All the targets are evolved stars: M67-141 is in
the red clump \citep{Yong} and the Pal 1 stars are all presumably on
the RGB.  If we assume that there has been minimal mass loss since the
stars evolved off the main sequence, and if we assume that not much
time has passed since the stars left the main sequence, then the main
sequence turn off mass can be used as the RGB stellar mass.  For most
Galactic GCs this turnoff mass is typically $0.8$
M$_{\sun}$---however, as discussed earlier, both M67 (an open cluster)
and Pal 1 are younger than the typical GC, and should therefore have a
higher turnoff mass than $0.8$ M$_{\sun}$.  To determine the turnoff
mass, we examined the Pal 1 isochrones computed by \citet{Sarajedini};
an age of $5\pm 1$ Gyr gives a turnoff mass of $M_{\rm{TO}} = 1.14 \pm
0.06$ M$_\sun$. This estimate agrees with that from \citet{Cohen} for
Pal 12, another young globular cluster.  M67-141 has an age from 4-5
Gyr from isochrone fits \citep{VandenbergStetson}, which means the
turnoff mass will be similar.  The final photometric gravities are
also shown in Table \ref{table:AtmParams}.

\begin{center}
\begin{deluxetable*}{lcccccc}
\centering
\tabletypesize{\scriptsize}
\tablecolumns{7}
\tablewidth{0pc}
\tablecaption{Atmospheric Parameters for the Pal 1 stars and the standard star, M67-141.\label{table:AtmParams}}
\tablehead{
 & \multicolumn{1}{l}{Photometric} & \multicolumn{2}{l}{Spectroscopic} \\
Star & $T_{\rm{eff}}$ (K) & $T_{\rm{eff}}$ (K) & $\log g$ & $\xi$ (km
s$^{-1}$) & [Fe/H] & $v\sin i$ (km s$^{-1}$)\tablenotemark{a}} 
\startdata
M67-141: This paper & $4624 \pm 12$ & $4750\pm 50$ & $2.31 \pm 0.06$ & 
$1.4\pm 0.1$ & $0.02\pm 0.02$ & 4.5 \\
\citet{Yong} & 4700 (4604)\tablenotemark{b} & 4700 & 2.3 & 1.34 & 0.00 & -\\
\citet{Pancino} & 4590 & 4650 & 2.8 & 1.3 & 0.06 & - \\
 & & & & & & \\
Monaco et al: Pal 1-I & 4850 & 5000 & 2.40 & 1.0 & -0.5 & - \\
Pal 1-I & $4742\pm 81$ & $4800\pm 70$ & $2.27\pm 0.15$ & $1.4\pm 0.3$
& $-0.61\pm 0.08$ & 2.5 \\
Pal 1-II & $4725\pm 80$ & $4750\pm 135$ & $2.33\pm 0.15$ & $1.6\pm
0.2$ & $-0.61\pm 0.08$ & 2.5 \\
Pal 1-III & $5061\pm 90$ & $5050\pm 50$ & $2.89\pm 0.15$ & $0.9\pm
0.1$ & $-0.60\pm 0.02$ & 0.5 \\
Pal 1-C & $4698\pm 79$ & $4750\pm 70$ & $2.22\pm 0.15$ & $1.3\pm 0.1$
& $-0.58\pm 0.03$ & 1.5 \\
\enddata
\tablenotetext{a}{This parameter was determined for spectrum sytheses,
as discussed in Section \ref{subsec:SS}.}
\tablenotetext{b}{The temperature in parentheses is the photometric
temperature with the spectroscopic metallicity.\\}
\end{deluxetable*} 
\end{center}

\subsection{Spectroscopic Parameters}\label{subsec:Spectroscopic}
Spectroscopic indicators are examined to refine these photometric
parameters.  The abundances of the individual Fe lines are analyzed in
MOOG \citep{Sneden},\footnote{The 2010 version of MOOG was obtained at
\url{http://www.as.utexas.edu/~chris/moog.html}.  We note that the
scattering version (MOOG-SCAT) was not used in this analysis because
of the high metallicities of our targets.    A test of MOOG-SCAT
showed that our line abundances would have negligible corrections: the
average correction for Pal 1-III is $\le$0.01 dex for lines below 5000
\AA.} using LTE OSMARCS model atmospheres with spherical geometries
(\citealt{MARCS}, \citealt{Plez}).  The final temperature was found by
forcing the \ion{Fe}{1} line abundances to be independent of the
excitation potential ($\chi$); similarly, the microturbulence was
refined by forcing the \ion{Fe}{1} abundances to be independent of the
line equivalent width.  The associated errors in these parameters were
found using the $1\sigma$ errors of the slopes of the \ion{Fe}{1}
abundances versus $\chi$ and EW.

It is not clear whether spectroscopic methods are appropriate for
determining the gravity (i.e. ensuring that the \ion{Fe}{1} and
\ion{Fe}{2} abundances are equal), since \ion{Fe}{1} is expected to
suffer from non-LTE effects which depopulate this under represented
species through interactions with the radiation field.  In addition,
there are few \ion{Fe}{2} lines in our spectral range; those that are
available are located at blue wavelengths so that reliable \ion{Fe}{2}
abundance determinations can be difficult, especially in low SNR
spectra.  For these reasons, we choose to adopt the photometric
surface gravities and their associated errors.

The final metallicity is the average \ion{Fe}{1} abundance, with an
error determined from the line-to-line scatter,
$\sigma(\textrm{\ion{Fe}{1}})$, divided by the square root of the
number of lines ($\delta(\textrm{\ion{Fe}{1}}) =
\sigma(\textrm{\ion{Fe}{1}})/\sqrt{N}$; see Section
\ref{subsubsec:statistical}).  The final spectroscopic values
($T_{\rm{eff}}$, $\log g$, $\xi$, and [Fe/H]) for M67-141 and the Pal
1 stars are listed in Table \ref{table:AtmParams}.  The spectroscopic
and photometric temperatures are in excellent agreement.  We also list
the atmospheric parameters from the literature for M67-141 (from
\citealt{Yong} and \citealt{Pancino}) and for Pal 1-I (from
\citealt{MonacoPal1}).  Our values for M67-141 are in excellent
agreement with \citet{Yong}, but our temperature and gravity disagree
with \citet{Pancino}.  Our Pal 1-I temperatures are slightly lower
and our microturbulence values are slightly higher than
\citet{MonacoPal1}. The differences are primarily due to differing
atomic data; as we explain in Section \ref{subsubsec:Monaco}, these
atmospheric parameter differences result in slightly different
abundances in the various analyses.

\begin{center}
\begin{figure}
\epsscale{1.25}
\plotone{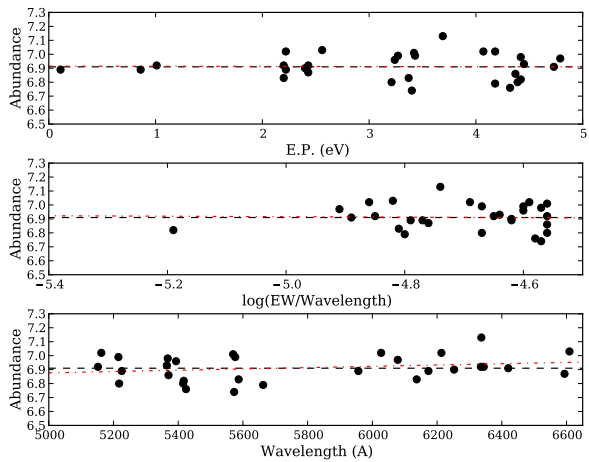}
\caption{The abundance trends with excitation potential (top), reduced
EW (middle) and wavelength (bottom) for Pal 1-I, with the spectroscopic
  atmospheric parameters listed in Table \ref{table:AtmParams}.  The
black dashed line shows a constant trend, while the red dot-dashed
line shows the trends of the data; note that in all three panels the
trend is essentially constant, though there is still a significant
dispersion.\\ \label{fig:MOOG}}
\end{figure}
\end{center}

Figure \ref{fig:MOOG} shows the abundances from the thirty \ion{Fe}{1}
lines that were measured in Pal 1-I, generated using the final
spectroscopic atmospheric parameters.  Note that though there is
little to no trend in abundance for excitation potential, reduced
equivalent width, or wavelength, there is still a range of $\sim 0.2$
dex.

\section{Elemental Abundances}\label{sec:Abundances}
The elemental abundances were determined in MOOG using EW matching.
With these abundances, the logarithmic ratios with respect to Solar
values, [X/Fe], are calculated.  Tables \ref{table:M67Abundances} and
\ref{table:Abundances} show the [X/Fe] ratios, the statistical errors
(see Section \ref{subsec:errors} for a description of how these errors
are determined), the number of lines used to determine the abundances,
and comparisons with literature values. These results will be
discussed in Section \ref{sec:Results}.

\begin{center}
\begin{deluxetable*}{lccccccc}
\tabletypesize{\scriptsize}
\tablecolumns{8}
\tablewidth{0pc}
\tablecaption{A comparison of the derived abundances in the standard
star, M67-141 from this work, \citet{Yong}, and \citet{Pancino}.  We
also give the total error in [X/Fe] and the number of lines used in
the calculation.\label{table:M67Abundances}}
\tablehead{
 & \multicolumn{2}{l}{\citet{Pancino}} & \multicolumn{2}{l}{\citet{Yong}} & \multicolumn{3}{l}{This work}\\
X & [X/Fe]\tablenotemark{a} & $\delta$ & [X/Fe]\tablenotemark{a} & $\delta$ & [X/Fe]\tablenotemark{a} & $\delta$ & N}
\startdata
\ion{Fe}{1} & 0.06 & 0.01 & -0.01 & 0.12 & 0.02 & 0.02 & 91 \\
\ion{Fe}{2} & 0.01 & 0.03 & 0.01 & 0.09 & -0.03 & 0.04 & 24 \\
\ion{O}{1} & -0.05 & 0.09 & 0.10 & 0.06 & -0.15 (0.03) & 0.09 & 2 \\
\ion{Na}{1} & 0.10 & 0.02 & 0.24 & 0.10 & 0.29 & 0.13 & 4 \\
\ion{Mg}{1} & 0.29 & 0.03 & 0.18 & 0.04 & 0.04 & 0.07 & 3 \\
\ion{Al}{1} & 0.06 & 0.06 & 0.16 & 0.06 & 0.14 & 0.15 & 2 \\
\ion{Si}{1} & 0.09 & 0.02 & 0.11 & 0.08 & 0.13 & 0.05 & 10 \\
\ion{Ca}{1} & -0.13 & 0.02 & 0.09 & 0.03 & 0.07 & 0.06 & 15\\
\ion{Sc}{2} & -0.02 & 0.08 & - & - & 0.09 & 0.06 & 12 \\
\ion{Ti}{1} & -0.07 & 0.02 & 0.05 & 0.05 & 0.02 & 0.05 & 35 \\
\ion{Ti}{2} & -0.07 & 0.02 & - & - & 0.10 & 0.05 & 33 \\
\ion{V}{1} & 0.13 & 0.04 & - & - & -0.05 & 0.05 & 17 \\
\ion{Cr}{1} & 0.01 & 0.03 & - & - & -0.04 & 0.09 & 8 \\
\ion{Cr}{2} & - & - & - & - & 0.18 & 0.18 & 2 \\
\ion{Mn}{1} & - & - & -0.20 & 0.03 & -0.08 & 0.13 & 4 \\
\ion{Co}{1} & 0.11 & 0.02 & 0.01 & 0.09 & -0.15 & 0.21 & 2\\
\ion{Ni}{1} & 0.06 & 0.05 & 0.06 & 0.09 & 0.08 & 0.04 & 26 \\
\ion{Cu}{1} & - & - & - & - & 0.16 & 0.13 & 3 \\
\ion{Zn}{1} & - & - & - & - & 0.05 & 0.12 & 3\\
\ion{Y}{2} & -0.04 & 0.02 & - & - & -0.04 & 0.11 & 5 \\
\ion{Ba}{2} & 0.26 & 0.05 & 0.02 & - & 0.0 & 0.17 & 2 \\
\ion{La}{2} & 0.06 & 0.05 & 0.13 & 0.04 & 0.04 & 0.06 & 4 \\
\ion{Nd}{2} & 0.01 & 0.29 & - & - & 0.04 & 0.08 & 5 \\
\ion{Eu}{2} & - & - & 0.05 & - & 0.0 & 0.12 & 1\\
\enddata
\tablenotetext{a}{[X/H] is given instead for \ion{Fe}{1} and
\ion{Fe}{2}.}
\end{deluxetable*}
\end{center}

\begin{center}
\begin{deluxetable*}{lcccccccccccccccccc}
\centering
\tabletypesize{\scriptsize}
\tablecolumns{14}
\tablewidth{0pc}
\tablecaption{[X/Fe] ratios, $1\delta$ errors, and the number of lines
used, for the Pal 1 stars.  The [X/Fe] ratios from \citet{MonacoPal1}
are also shown for comparison.\label{table:Abundances}}
\tablehead{
 & Monaco Pal 1-I &\multicolumn{3}{l}{Pal 1-I} & \multicolumn{3}{l}{Pal 1-II} &
  \multicolumn{3}{l}{Pal 1-III} & \multicolumn{3}{l}{Pal 1-C}\\
X & [X/Fe]\tablenotemark{a} & [X/Fe]\tablenotemark{a} & $\delta$ & $N$
& [X/Fe]\tablenotemark{a} & $\delta$ & $N$ & [X/Fe]\tablenotemark{a} &
$\delta$ & $N$ & [X/Fe]\tablenotemark{a} & $\delta$ & $N$}
\startdata
\ion{Fe}{1} & -0.49 & -0.61 & 0.08 & 30 & -0.61 & 0.08 & 21 & -0.60 & 0.03 & 81 & -0.58 & 0.03 & 60 \\
\ion{Fe}{2}  & -0.53 & -0.46 & 0.30 & 1 & -0.64 & 0.44 & 2 & -0.61 & 0.05 & 23 & -0.68 & 0.05 & 23 \\
\ion{O}{1} & - & $<0.82$ & - & 1 & $<0.32$ & - & 1 & $<0.42$ & - & 1 & 0.20 (0.23) & 0.11 & 2 \\
\ion{Na}{1} & 0.38 & 0.26 & 0.34 & 2 & 0.23 & 0.31 & 2 & 0.20 & 0.09 & 4 & 0.16 & 0.09 & 4 \\
\ion{Mg}{1} & 0.11 & -0.11 & 0.20 & 1 & -0.13 & 0.30 & 1 & -0.02 & 0.08 & 5 & 0.02 & 0.07 & 6 \\
\ion{Al}{1} & 0.25 & - & - & - & - & - & - & 0.0 & 0.10 & 2 & 0.07 & 0.10 & 2 \\
\ion{Si}{1} & -0.01 & 0.24 & 0.24 & 1 & 0.13 & 0.23 & 1 & 0.09 & 0.06 & 8 & 0.19 & 0.06 & 8 \\
\ion{Ca}{1} & 0.04 & 0.16 & 0.16 & 6 & -0.04 & 0.22 & 4 & 0.15 & 0.06 & 17 & 0.12 & 0.06 & 18 \\
\ion{Sc}{2} & -0.01 & 0.28 & 0.26 & 3 & 0.21 & 0.33 & 2 & 0.30 & 0.09 & 8 & 0.21 & 0.07 & 12 \\
\ion{Ti}{1} & 0.10 & -0.03 & 0.34 & 3 & -0.14 & 0.38 & 4 & 0.02 & 0.06 & 24 & -0.08 & 0.06 & 24 \\
\ion{Ti}{2} & - & - & - & - & -0.13 & 0.51 & 2 & -0.08 & 0.05 & 32 & 0.08 & 0.07 & 23 \\
\ion{V}{1} & -0.06 & -0.05 & 0.21 & 1 & 0.08 & 0.14 & 4 & 0.14 & 0.07 & 6 & 0.06 & 0.05 & 14 \\
\ion{Cr}{1} & -0.23 & - & - & - & - & - & - & 0.01 & 0.09 & 11 & -0.16 & 0.11 & 9 \\
\ion{Cr}{2} & - & - & - & - & - & - & - & 0.17 & 0.20 & 2 & 0.19 & 0.21 & 2 \\
\ion{Mn}{1} & -0.22 & - & - & - & -0.16 & 0.36 & 1 & -0.21 & 0.12 & 4 & -0.11 & 0.10 & 8 \\
\ion{Co}{1} & -0.06 & - & - & - & - & - & - & -0.10 & 0.18 & 2& 0.03 & 0.20 & 2 \\
\ion{Ni}{1} & -0.03 & 0.05 & 0.25 & 3 & 0.09 & 0.14 & 5 & 0.08 & 0.05 & 20 & 0.03 & 0.05 & 21 \\
\ion{Cu}{1} & - & - & - & - & -0.05 & 0.33 & 1 & 0.04 & 0.18 & 2 & -0.08 & 0.32 & 1 \\
\ion{Zn}{1} & 0.38 & - & - & - & - & - & - & 0.08 & 0.14 & 3 & 0.15 & 0.15 & 3 \\
\ion{Y}{2} & -0.32 & $<0.04$ & - & 1 & $<0.34$ & - & 1 & -0.45 & 0.12 & 4 & -0.36 & 0.12 & 5\\
\ion{Ba}{2} & 0.24 & 0.27 & 0.24 & 2 & 0.19 & 0.26 & 2 & 0.22 & 0.12 & 3 & 0.26 & 0.18 & 2 \\
\ion{La}{2} & 0.29 & $<0.38$ & - & 1 & $<0.78$ & - & 1 & 0.42 & 0.16 & 1 & 0.24 & 0.14 & 3 \\
\ion{Nd}{2} & - & 0.19 & 0.65 & 1 & - & - & - & 0.13 & 0.16 & 2 & 0.12 & 0.14 & 2 \\
\ion{Eu}{2} & - & $<0.78$ & - & 1 & $<0.78$ & - & 1 & 0.50 & 0.20 & 1 & 0.50 & 0.20 & 1 \\
\enddata
\tablenotetext{a}{[X/H] is given instead for \ion{Fe}{1} and
\ion{Fe}{2}.}
\end{deluxetable*}
\end{center}

\vspace{-0.75in}

\subsection{Hyperfine Structure}\label{subsec:HFS}
The elements Sc, V, Mn, Co, Cu, Ba, La, and Eu were checked for
hyperfine structure (HFS) corrections.  The majority of the HFS and
isotopic data comes from \citet[for Sc, V, Mn, and Co]{Prochaska},
\citet[additional Mn]{Booth}, \citet[Cu]{Biehl}, \citet[Ba]{McW98},
and \citet[La and Eu]{LawlerA,LawlerB}, with extra lines added from
the Kurucz
database.\footnote{\url{http://kurucz.harvard.edu/linelists.html}}
The corrections per line were averaged over the entire star and were
applied to the average abundances from MOOG.  If HFS data were not
available, then the line was neglected in the correction calculation.
When the corrections were less than 0.1 dex they were ignored.
Negligible corrections were found ($<0.1$ dex) for Sc, Ba, La, and Eu;
moderate corrections ($\lesssim 0.5$ dex) for V and Mn; and large
corrections ($\lesssim 1.0$ dex) for Co and Cu.

\subsection{Spectrum Synthesis}\label{subsec:SS}
When there was only a single absorption line for an important element,
a spectrum synthesis was performed on the line.  In each spectrum
synthesis the instrumental and rotational broadening are taken into
account.  With resolutions of $R=36000$ and $R=60000$, Pal 1 and
M67-141 have instrumental broadening values of 8.3 and 5.0 km
s$^{-1}$, respectively.  The rotational broadenings for each star were
determined by examining nearby lines with known abundances, and
adjusting the $v \sin i$ values until the widths and depths matched
the points. For example, in the case of the the \ion{Eu}{2} 6643 \AA
\hspace{0.025in} line, the 6644 \AA \hspace{0.025in} \ion{Ni}{1} line
and the 6647 \AA \hspace{0.025in} \ion{Fe}{1} line were used to
determine these parameters.  The derived $v\sin i$ values for M67-141
and the four Pal 1 stars are shown in Table \ref{table:AtmParams}.
The abundances were then estimated from the best fits to the adopted
models, with an error in $\log \epsilon$ equal to the range of
abundances that fit the line profiles.  Where the spectral lines were
particularly noisy, only upper limits are provided.  Spectrum
syntheses were also used when a known blend affected an important
line, as in the case of the 6141.73 \AA \hspace{0.025in} \ion{Ba}{2}
line, which is blended with a \ion{Ni}{1} line at solar metallicity
\citep{AllendePrieto2001}.  By using a spectrum synthesis the strength
of the Ni line could also be taken into account.

\begin{figure}
\epsscale{1.25}
\plotone{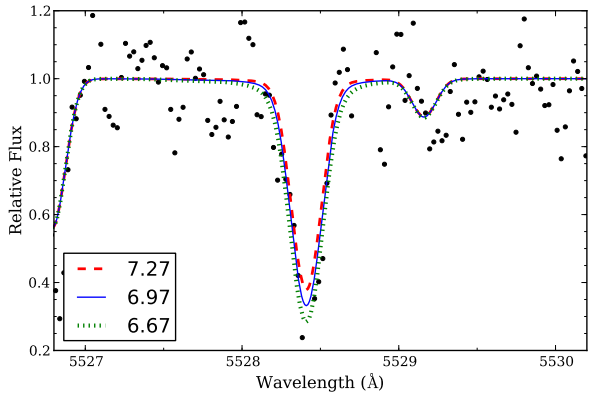}
\caption{Spectrum synthesis for the \ion{Mg}{1} 5528 \AA
\hspace{0.025in} line in Pal 1-I  Syntheses for three different Mg
abundances are shown: $\log \epsilon(\rm{Mg}) = 7.27$, $6.97$, and
$6.67$.  The parameters for the synthesis are discussed in Section
\ref{subsec:SS}.  Note that though the fits match the width of the
line well, the bottom-most point is not fit; this point is most likely
a noise spike.\\ \label{fig:SS}}
\end{figure}

An example of a spectrum synthesis of the \ion{Mg}{1} 5528 \AA
\hspace{0.04in} line in Pal 1-I is shown in Figure \ref{fig:SS}.
Note that the width of the line is fit well, but the bottom
point is missed; we suspect that this bottom point is a noise spike.

\subsection{Errors}\label{subsec:errors}
To determine the errors on the abundances, we separate systematic
errors (due to model atmosphere parameters) from statistical errors
(due to measurements).  While the systematic errors are calculated and
discussed, they are not included in the errors bars in our figures and
are tabulated only in Tables \ref{table:AtmErrorsM67-141} and
\ref{table:AtmErrorsPal3}.

\subsubsection{Systematic Errors}\label{subsubsec:systematic}
The abundances from individual spectral lines in a star are all
affected by the choice of model atmosphere parameters.  Changing each
parameter ($T_{\rm{eff}}$, $\log g$, $\xi$, [M/H]) will affect the
average abundance of each species.  While the errors in the
atmospheric parameters themselves are random, changes in these
parameters will systematically affect the abundance analyses.

To determine how the abundances were affected by uncertainties in the
atmospheric parameters, new model atmospheres were tested by varying
one parameter at a time by its $1\sigma$ error.  These are quoted
individually in Tables \ref{table:AtmErrorsM67-141} and
\ref{table:AtmErrorsPal3}.  It is not clear how to combine these
errors, since they are not independent, e.g. increasing $\log g$
affects the $T_{\rm{eff}}$ and $\xi$ estimates in our analysis.
However, investigating their effects separately provides a maximum
uncertainty estimate.  We also consider systematic errors due to
continuum placement; as discussed in Section \ref{sec:EWs}, continuum
placement can result in EW differences that are outside our adopted
$1\sigma$ errors.  As a conservative estimate of these errors, we
adopt the average offset between our EW measurements and those of
\citet[4 m\AA, for the high SNR stars]{Pancino} and
\citet[7 m\AA, for the low SNR stars]{MonacoPal1}, for our continuum
uncertainty.

Representative systematic errors in abundances for M67-141, Pal 1-I,
and Pal 1-III are shown in Tables \ref{table:AtmErrorsM67-141} and
\ref{table:AtmErrorsPal3}, individually and added in quadrature.  The
differences are largest for the continuum and temperature errors in
M67-141, while the errors are fairly similar for all parameters in the
Pal 1 stars.  The final errors in the [X/Fe] ratios are fairly small
for all elements.
\begin{center}
\begin{deluxetable*}{lcccccc}
\centering
\tabletypesize{\scriptsize}
\tablecolumns{7}
\tablewidth{0pc}
\tablecaption{Errors in M67-141's [X/Fe] ratios due to atmospheric
parameter errors.  The total error shows the individual errors added
in quadrature.\label{table:AtmErrorsM67-141}}
\tablehead{
 & $\Delta T=+50$ K & $\Delta \log g=+0.06 $ & $\Delta \xi = +0.1$ km
s$^{-1}$ & $\Delta [\rm{M/H}] = +0.02$ & Continuum +4 m\AA & Total}
\startdata
$\Delta$[\ion{Fe}{1}/H] & +0.07 & +0.01 & -0.05 & +0.01 & +0.07 & 0.11\\
$\Delta$[\ion{Fe}{2}/H] & -0.07 & +0.03 & -0.05 & +0.01 & +0.09 & 0.13 \\
$\Delta$[\ion{O}{1}/Fe] & -0.06 & +0.02 & +0.05 & 0.0 & +0.02 & 0.08\\
$\Delta$[\ion{Na}{1}/Fe] & +0.01 & -0.01 & +0.02 & -0.01 & -0.01 & 0.03\\
$\Delta$[\ion{Mg}{1}/Fe] & -0.03 & 0.0 & +0.04 & 0.0 & 0.0 & 0.05 \\
$\Delta$[\ion{Al}{1}/Fe] & 0.0 & -0.01 & +0.03 & -0.01 & -0.01 & 0.03\\
$\Delta$[\ion{Si}{1}/Fe] & -0.10 & 0.0 & +0.03 & -0.01 & 0.0 & 0.10 \\
$\Delta$[\ion{Ca}{1}/Fe] & +0.04 & -0.01 & 0.0 & -0.01 & -0.01 & 0.04\\
$\Delta$[\ion{Sc}{2}/Fe] & -0.08 & +0.01 & 0.0 & 0.0 & 0.0 & 0.08\\
$\Delta$[\ion{Ti}{1}/Fe] & +0.09 & -0.01 & -0.01 & -0.01 & 0.0 & 0.09\\
$\Delta$[\ion{Ti}{2}/Fe] & -0.08 & +0.02 & -0.01 & 0.0 & +0.02 & 0.09\\
$\Delta$[\ion{V}{1}/Fe] & +0.11 & 0.0 & +0.01 & 0.0 & +0.01 & 0.11\\
$\Delta$[\ion{Cr}{1}/Fe] & +0.08 & -0.01 & -0.02 & -0.01 & 0.01 & 0.08\\
$\Delta$[\ion{Cr}{2}/Fe] & -0.13 & +0.01 & -0.01 & -0.01 & +0.02 & 0.13\\
$\Delta$[\ion{Mn}{1}/Fe] & +0.05 & 0.0 & -0.02 & 0.0 & 0.0 & 0.05\\
$\Delta$[\ion{Co}{1}/Fe] & 0.0 & 0.0 & -0.02 & -0.01 & +0.01 & 0.02\\
$\Delta$[\ion{Ni}{1}/Fe] & -0.03 & 0.0 & 0.0 & -0.01 & 0.0 & 0.03\\
$\Delta$[\ion{Cu}{1}/Fe] & 0.0 & -0.01 & -0.01 & 0.0 & -0.02 & 0.02\\
$\Delta$[\ion{Zn}{1}/Fe] & -0.11 & +0.01 & 0.0 & 0.0 & +0.03 & 0.11\\
$\Delta$[\ion{Y}{2}/Fe] & -0.07 & +0.02 & -0.02 & 0.0 & +0.03 & 0.08\\
$\Delta$[\ion{Ba}{2}/Fe] & -0.04 & +0.02 & -0.03 & +0.01 & 0.0 & 0.05\\
$\Delta$[\ion{La}{2}/Fe] & -0.05 & +0.02 & +0.04 & 0.0 & +0.02 & 0.07\\
$\Delta$[\ion{Nd}{2}/Fe] & -0.06 & +0.02 & +0.02 & 0.0 & +0.05 & 0.08\\
$\Delta$[\ion{Eu}{2}/Fe] & -0.08 & +0.02 & +0.04 & 0.0 & +0.02 & 0.09\\
\enddata
\end{deluxetable*} 
\end{center}

\begin{center}
\begin{deluxetable*}{lcccccccccccc}
\centering
\tabletypesize{\scriptsize}
\tablecolumns{13}
\tablewidth{0pc}
\tablecaption{Errors in Pal 1-I and -III's [X/Fe] ratios due to the atmospheric parameter errors.  The total error shows the individual errors added in quadrature.\label{table:AtmErrorsPal3}}
\tablehead{
 & \multicolumn{2}{c}{$\Delta T_{\rm{eff}}$ (K)} & \multicolumn{2}{c}{$\Delta \log g$} & \multicolumn{2}{c}{$\Delta \xi$ (km s$^{-1}$)} & \multicolumn{2}{c}{$\Delta$ [M/H]} & \multicolumn{2}{c}{Continuum (m\AA)} & \multicolumn{2}{c}{Total} \\
 & I & III & I & III & I & III & I & III & I & III & I & III \\
 & $+70$ & $+50$ & $+0.15 $ & $+0.15 $ & $+0.3$ & $+0.1$ & $+0.08$ & $+0.02$ & $+7$ & $+4$ & & }
\startdata
$\Delta$[\ion{Fe}{1}/H] & +0.07 & +0.05 & 0.0 & -0.01 & -0.18 & -0.04 & +0.01 & 0.0 & +0.12 & +0.08 & 0.23 & 0.10 \\
$\Delta$[\ion{Fe}{2}/H] & -0.04 & -0.03 & +0.05 & +0.06 & -0.08 & -0.04 & +0.02 & 0.0 & +0.11 & +0.14 & 0.15 & 0.16 \\
$\Delta$[\ion{O}{1}/Fe] & - & -0.04 & - & +0.08 & - & +0.04 & - & +0.01 & - & +0.14 & - & 0.17\\
$\Delta$[\ion{Na}{1}/Fe] & -0.01 & -0.02 & -0.02 & -0.01 & +0.08 & +0.02 & -0.01 & 0.0 & -0.01 & -0.03 & 0.08 & 0.04\\
$\Delta$[\ion{Mg}{1}/Fe] & -0.02 & -0.02 & -0.05 & -0.02 & +0.08 & +0.03 & 0.0 & 0.0 & -0.06 & -0.02 & 0.11 & 0.05 \\
$\Delta$[\ion{Al}{1}/Fe] & - & -0.02 & - & +0.01 & - & +0.04 & - & 0.01 & - & +0.03 & - & 0.05\\
$\Delta$[\ion{Si}{1}/Fe] & -0.08 & -0.04 & +0.02 & +0.03 & +0.12 & +0.04 & 0.0 & +0.01 & -0.01 & +0.02 & 0.15 & 0.07\\
$\Delta$[\ion{Ca}{1}/Fe] & 0.0 & 0.0 & -0.02 & -0.02 & +0.03 & +0.01 & -0.01 & 0.0 & -0.02 & -0.04 & 0.04 & 0.05\\
$\Delta$[\ion{Sc}{2}/Fe] & -0.08 & -0.05 & +0.06 & +0.07 & +0.09 & 0.0 & +0.01 & 0.0 & +0.01 & +0.05 & 0.14 & 0.10\\
$\Delta$[\ion{Ti}{1}/Fe] & +0.04 & +0.02 & +0.01 & +0.01 & 0.0 & 0.0 & -0.01 & 0.0 & +0.04 & +0.02 & 0.06 & 0.03\\
$\Delta$[\ion{Ti}{2}/Fe] & - & -0.05 & - & +0.06 & - & 0.0 & - & 0.0 & - & +0.05 & - & 0.09\\
$\Delta$[\ion{V}{1}/Fe]  & +0.05 & +0.02 & 0.0 & +0.02 & +0.14 & +0.03 & -0.01 & 0.0 & 0.0 & +0.03 & 0.15 & 0.05\\
$\Delta$[\ion{Cr}{1}/Fe] & - & +0.02 & - & -0.01 & - & -0.01 & - & 0.0 & - & -0.01 & - & 0.03\\
$\Delta$[\ion{Cr}{2}/Fe] & - & -0.07 & - & +0.07 & - & 0.0 & - & +0.01 & - & +0.07 & - & 0.12\\
$\Delta$[\ion{Mn}{1}/Fe] & - & +0.01 & - & -0.01 & - & 0.0 & - & +0.01 & - & -0.02 & - & 0.03\\
$\Delta$[\ion{Co}{1}/Fe] & - & 0.0   & - & +0.03 & - & +0.02 & - & +0.01 & - & +0.04 & - & 0.05\\
$\Delta$[\ion{Ni}{1}/Fe] & -0.01 & -0.02 & +0.02 & +0.01 & +0.01 & 0.0 & -0.01 & 0.0 & +0.01 & +0.01 & 0.03 & 0.02\\
$\Delta$[\ion{Cu}{1}/Fe] & - & 0.0   & - & +0.01 & - & -0.02 & - & 0.0 & - & +0.01 & - & 0.02\\
$\Delta$[\ion{Zn}{1}/Fe] & - & -0.07 & - & +0.04 & - & 0.0 & - & 0.0 & - & +0.05 & - & 0.09\\
$\Delta$[\ion{Y}{2}/Fe]  & - & -0.05 & - & +0.07 & - & 0.0 & - & +0.01 & - & +0.08 & - & 0.12\\
$\Delta$[\ion{Ba}{2}/Fe] & -0.05 & -0.04 & +0.05 & +0.04 & -0.08 & -0.02 & +0.02 & +0.01 & 0.0 & -0.01 & 0.11 & 0.06\\
$\Delta$[\ion{La}{2}/Fe] & - & -0.04 & - & +0.07 & - & +0.03 & - & 0.0 & - & +0.10 & - & 0.13\\
$\Delta$[\ion{Nd}{2}/Fe] & -0.05 & -0.05 & +0.07 & +0.07 & +0.09 & +0.01 & +0.01 & 0.0 & +0.05 & +0.08 & 0.13 & 0.12\\
$\Delta$[\ion{Eu}{2}/Fe] & - & -0.05 & - & +0.08 & - & +0.04 & - & +0.01 & - & +0.14 & - & 0.17\\
\enddata
\end{deluxetable*} 
\end{center}

\vspace{-0.4in}

\subsubsection{Statistical Errors}\label{subsubsec:statistical}
Following the method outlined by \citet{Shetrone}, we identify three
types of statistical errors for the abundances: the error from the
\ion{Fe}{1} line-to-line scatter, $\sigma$(Fe); the error from the
line-to-line scatter of the element itself, $\sigma$(X); and the error
in the abundances due to uncertainty in the measured equivalent
widths, $\sigma$(EW). Because there are quite a few \ion{Fe}{1} lines,
many of them strong enough to be detectable in the low SNR spectra,
$\sigma$(Fe) should provide a  conservative minimum estimate of the
line-to-line scatter due to the SNR and continuum placement.  If
$\sigma$(X) is less than $\sigma$(Fe), we assume that $\sigma$ has
been underestimated because the spectrum has too few lines measurable
lines for element X.  To estimate the errors due to EW measurements,
the \citet{Cayrel} formula is used to determine the minimum EW
measurement error ($\Delta$EW$_{\rm{min}}$) given the SNR and
resolution per spectrum; the EW error for a single line is then
$\Delta(\rm{EW}) = \Delta(\rm{EW}_{\rm{min}}) + 10\% \; \rm{EW}$ (also
see \citealt{Shetrone}).  These EW errors are propagated through the
abundance analyses, providing an average $\sigma$(EW).  For each
element, the larger of these three errors ($\sigma$(Fe), $\sigma$(X),
$\sigma$(EW)) is selected, and divided by $\sqrt{N_{\rm{lines}}}$
(where $N_{\rm{lines}}$ is the number of lines of that element) to
determine the total statistical error, $\delta$ (i.e. $\delta =
\sigma/\sqrt{N}$). The total $1\delta$ errors in [Fe/H] and [X/Fe] are
listed in Tables \ref{table:M67Abundances} and \ref{table:Abundances}.

\subsection{Solar Abundances}\label{subsec:solar}
An analysis was also performed of the solar spectrum from the Kurucz
2005 solar flux
atlas.\footnote{\url{http://kurucz.harvard.edu/sun.html}} The model
atmosphere was generated with $T_{\rm{eff}} = 5777$ K, $\log g = 4.44$
dex, $\xi = 0.85$ km s$^{-1}$, and $[M/H] = 0.0$ \citep{Yong}.  Lines
were discarded if they were too weak, too strong, or looked to be an
unresolved blend at the solar $T_{\rm{eff}}$ and [Fe/H].  The final
solar abundances are shown in Table \ref{table:Sun} along with the
\citet{Asplund} values for reference.  The results are in excellent
agreement, though O, Sc, and Ba are all higher in our analysis, a
trend that was also noticed by \citet[for O and Ba]{Yong}, \citet[for
Ba]{Pancino} and \citet[for Sc and Ba]{MonacoPal1}.  By using these
high solar values, systematically high [O/Fe], [Sc/Fe], and [Ba/Fe]
values will be removed from our Pal 1 analysis.  Because the single Eu
line is very weak in the solar spectrum, we simply use the
\citet{Asplund} value.  Our solar values are adopted in all
calculations of [Fe/H] and [X/Fe].

\begin{center}
\begin{deluxetable}{lll}
\centering
\tabletypesize{\scriptsize}
\tablecolumns{3}
\tablewidth{0pc}
\tablecaption{Our solar abundances compared to those from
\citet{Asplund}.\label{table:Sun}}
\tablehead{
Element & Asplund et al. (2009) & This work}
\startdata
\ion{Fe}{1} & 7.45 & 7.50 \\
\ion{Fe}{2} & 7.45 & 7.49 \\
\ion{O}{1} & 8.69 & 8.83 (8.85)\tablenotemark{a} \\
\ion{Na}{1} & 6.27 & 6.25 \\
\ion{Mg}{1} & 7.53 & 7.59 \\
\ion{Al}{1} & 6.43 & 6.47 \\
\ion{Si}{1} & 7.51 & 7.51 \\
\ion{Ca}{1} & 6.29 & 6.32 \\
\ion{Sc}{2} & 3.05 & 3.19 \\
\ion{Ti}{1} & 4.91 & 4.93 \\
\ion{Ti}{2} & 4.91 & 4.95 \\
\ion{V}{1} & 3.96 & 3.95 \\
\ion{Cr}{1} & 5.64 & 5.63 \\
\ion{Cr}{2} & 5.64 & 5.59 \\
\ion{Mn}{1} & 5.48 & 5.41\tablenotemark{b} \\
\ion{Co}{1} & 4.87 & 4.91\tablenotemark{b} \\
\ion{Ni}{1} & 6.20 & 6.27 \\
\ion{Cu}{1} & 4.25 & 4.28\tablenotemark{b} \\
\ion{Zn}{1} & 4.63 & 4.58 \\
\ion{Y}{2} & 2.17 & 2.18 \\
\ion{Ba}{2} & 2.18 & 2.29 \\
\ion{La}{2} & 1.17 & 1.22 \\
\ion{Nd}{2} & 1.45 & 1.55 \\
\ion{Eu}{2} & 0.51 & 0.51\tablenotemark{c} \\
\enddata
\tablenotetext{a}{The number in parentheses is the O abundance as
determined with MOOG 2002.}
\tablenotetext{b}{With HFS corrections}
\tablenotetext{c}{From Asplund et al. (2009)\\}
\end{deluxetable}
\end{center}

\subsection{NLTE Effects}\label{subsec:NLTE}
Non-LTE corrections are neglected in this analysis, yet it is well
known that many of our elements do suffer from NLTE effects.  We have
tried to minimize this problem by eliminating lines which are greatly
affected by NLTE.   Na is a particularly sensitive element; however,
weaker lines tend to have smaller corrections.  At solar metallicity
\citet{Lind} recommend the 6154.23/6160.75 \AA \hspace{0.025in}
doublet, with the 5682.65/5688.21 \AA \hspace{0.025in} doublet for
$[\rm{Fe/H}]<-1.0$. For the Sun and M67-141 only the former doublet is
used; for Pal 1, however, both sets are used.  Any corrections to
these lines should be $< -0.10$ dex at most \citep{Mashonkina2000}. 

The NLTE effects for Mn can also be quite large.  However,
\citet{BergemannGehren} note that the corrections are strongest for
the weak and intermediate lines (i.e. those with EW $<$ 80 m\AA).  All
of our lines are stronger than 80 m\AA---such corrections are then
``scattered around zero or negative.''

Barium is another element that suffers from NLTE effects that can
broaden the lines.  The 6496.91 \AA \hspace{0.025in} line has been
eliminated from this analysis, since it has large corrections.
However the corrections for 5853.69 \AA \hspace{0.025in} and 6141.73
\AA \hspace{0.025in} are negligible for Pal 1's metallicity range, and
the 4554.03 \AA \hspace{0.025in} correction is small at solar
metallicity \citep{ShortHauschildt}.

\subsection{Comparisons with Previous Work}\label{subsec:comparison}

We compare our [X/Fe] results for M67-141 to \citet{Yong} and
\citet{Pancino}, and for Pal 1-I to \citet{MonacoPal1}, in Tables
\ref{table:M67Abundances} and \ref{table:Abundances}. Here we discuss
any discrepancies.

\vspace{0.4in}
\subsubsection{M67-141: Yong et al. (2005), Pancino et al. (2010)}\label{subsubsec:Yong}
Our [X/Fe] values agree very well with those of \citet[see Table
\ref{table:M67Abundances}; note that we do not shift their values to
common solar abundances as they derive their own solar
abundances]{Yong}.  The only inconsistent abundances are those of O
and Mg.  The Mg difference may be due to the choice of lines. The O
abundance, however, relies on molecular equilibrium calculations
within MOOG; these calculations seem to differ between the 2002
version used by Yong et al. and the 2010 version we use.  Using the
2002 version increases our O abundances and brings the results into
better agreement.  These abundances are shown in Tables
\ref{table:M67Abundances}, \ref{table:Abundances}, and
\ref{table:Sun}, in parentheses following the 2010 versions.  Note
that the solar abundance is only slightly affected.  As many of the
comparison studies in Section \ref{sec:Results} use the 2002 version
of MOOG, we consider that version to be more accurate for
comparisons.

Our [X/Fe] ratios do not agree as well with \citet{Pancino}: Na, Mg,
Ca, Ti, V, Co, and Ba all disagree.  For Na, Al, Ti, and Ba the
discrepancies seem to be due primarily to the different atmospheric
parameters.  \citet{Pancino} also do not appear to have applied HFS
corrections to V or Co, which could explain why their abundances are
much higher than ours.  Mg is again discrepant because of the choice
of lines and atomic data.  Ca is also quite low in their analysis,
also apparently due to atomic data.

Finally, we note that a recent spectral comparison by \citet{Onehag}
of a solar twin in M67 suggests that the entire cluster should have
roughly solar abundances, with a slightly elevated $[\rm{Fe/H}] \sim
0.02$ dex.  With the exceptions of Na and Si, all of our elements are
within 0.08 dex of solar [X/Fe].  Na could very well be elevated in
clump stars like M67-141; \citet{Tautvaisiene2000} suggested that
mixing could take place in open clusters after the He-flash, which
could bring up material from the Na-Ne cycle.  Alternatively this high
Na abundance could indicate a need for a negative NLTE correction.

\subsubsection{Pal 1-I: Monaco et al. (2010)}\label{subsubsec:Monaco}
As shown in Table \ref{table:Abundances}, Monaco et al.'s analysis
differs from ours for Mg, Al, Sc, and Ti.  The significantly
different atmospheric parameters affect most of these ratios,
particularly Al (since our Pal 1-I spectrum does not have a
sufficiently high SNR to determine the Al abundance, we compare their
Pal 1-I abundance to our Pal 1-III and -C values, which assumes there
is not an abundance spread).  In addition, different atomic data for
Mg, Al, and Ti causes additional discrepancies.  The \ion{Sc}{2}
abundance seems to differ due to the solar Sc abundances; the measured
solar EWs in \citet{MonacoPal1} are considerably higher than ours,
leading to a higher solar Sc abundance and therefore a lower [Sc/Fe]
for Pal 1-I.

The differences in atomic data are rather discouraging; however, we
note that while the choices of lines are broad, much of our atomic
data overlaps with that of \citet{Fulbright} and
\citet{Reddy2003,Reddy2006}, who provide the majority of the
metal-rich Galactic stars for the comparisons in our analysis.
Therefore we consider that our results are fairly robust for this
comparison.

\vspace{0.4in}
\section{Abundance Results}\label{sec:Results}
Figures \ref{fig:AlphaFe1} through \ref{fig:BaYBaEuEuAlphaFe} compare
the abundance trends of the Pal 1 stars (red star symbols) and M67-141
(red circle) to those of Galactic field stars (field stars; grey points, 
from the sources assembled by \citealt{Venn2004}, with additions from
\citealt{Reddy2006} and \citealt{Simmerer}), Galactic GCs (black solid
symbols, from the sources assembled in \citealt{Pritzl}, plus
\citealt{Sbordone2007} and \citealt{Cohen}), and old, metal-poor
Galactic OCs (open circles, from \citealt{Carraro2004},
\citealt{Yong}, and \citealt{Friel2010}).  The two old, most
metal-poor OCs are Berkeley 20 (Be 20) and Be 29; the latter has been
linked to the Sgr dSph based on its location, chemistry, and
kinematics \citep{CarraroBensby} and its location on the Sgr
age-metallicity relation \citep{ForbesBridges}.  The bulge and
thin/thick disk GCs (circles) are distinguished from the halo clusters
Pal 12 (plus signs) and Ter 7 (crosses), both of which are associated
with the Sgr dSph. The bulge and disk cluster abundances are averaged
over the entire cluster, while the individual stars in the halo
clusters are separate.  Those studies that do not perform their own
solar abundance analyses have been shifted to the \citet{Asplund}
abundances.

\subsection{Metallicity}\label{subsec:metallicity}
The derived metallicities are in excellent agreement with previous
studies.  For M67-141, $[\rm{Fe/H}]=0.02\pm 0.02$, which agrees with
both \citet{Yong} and \citet{Onehag}.  The average metallicity of the 
four Pal 1 stars, $[\rm{Fe/H}]=-0.60 \pm 0.01$, is in agreement with
its CaT metallicity \citep{RosenbergB} and its metallicity from
isochrone fits \citep{Sarajedini}.  As discussed by
\citet{MonacoPal1}, this average metallicity is high for a GC so far
from the Galactic center---the only metal-rich Galactic halo GCs are
Pal 12 and Ter 7, neither of which are actually Galactic.  However,
there are bulge and disk clusters in this metallicity regime,
including 47 Tuc.  This metallicity is also low for an OC, though Be
20 and 29 are nearly as metal-poor.

We further note that the [\ion{Fe}{1}/H] and [\ion{Fe}{2}/H] ratios
are in rough agreement, suggesting that our results would not differ
if spectroscopic gravities had been used (see Section
\ref{sec:ModelAtms}).

\subsection{$\alpha$-Elements}\label{subsec:alpha}
\begin{figure}
\epsscale{1.25}
\plotone{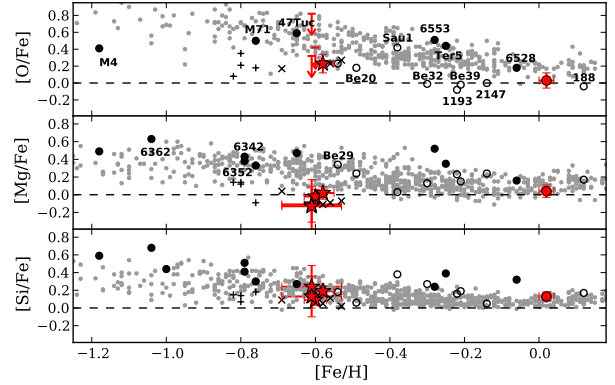}
\caption{[X/Fe] ratios of $\alpha$-elements O, Mg, and Si versus
metallicity for the stars in our study as well as Galactic field stars
and clusters.  The red star symbols are the four Pal 1 stars.  M67-141
is the red circle.  Points with arrows represent upper (or lower)
limits.  The grey points are Galactic field stars, from the sources
assembled by \citet{Venn2004} with additions from \citet{Reddy2006}.
Black open circles are Galactic OCs from \citet{Carraro2004},
\citet{Yong}, and \citet{Friel2010}. Black filled circles are bulge
and disk GCs from the sources assembled by \citet{Pritzl}. Plus signs
are stars in Palomar 12 (from \citealt{Cohen}) while crosses are stars
in Ter 7 (from \citealt{Sbordone2007}); both are halo GCs.  The dashed
lines show solar values.\label{fig:AlphaFe1}}
\end{figure}

\begin{figure}
\epsscale{1.25}
\plotone{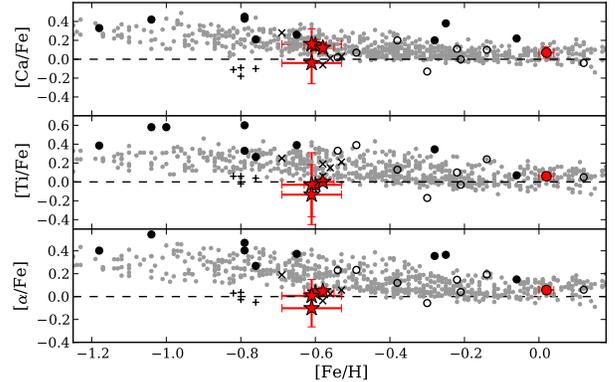}
\caption{[X/Fe] ratios of the $\alpha$-elements Ca and Ti, as
well as of the average $\alpha$ from Mg, Ca, and Ti.  The Ti abundance
is an average of \ion{Ti}{1} and \ion{Ti}{2}, except for Pal 1-I,
whose \ion{Ti}{1} abundance is shown since its \ion{Ti}{2} abundance
is abnormally high.  The points are as in Figure
\ref{fig:AlphaFe1}.\\ \label{fig:AlphaFe2}}
\end{figure}

The $\alpha$-elements form via the capture of $^{4}$He nuclei.  This
happens primarily during hydrostatic burning in massive stars, and the
$\alpha$-elements are subsequently released into the interstellar
medium by Type II supernovae.  Figures \ref{fig:AlphaFe1} and
\ref{fig:AlphaFe2} show the [X/Fe] ratios versus [Fe/H] for
$\alpha$-elements (O, Mg, Si, and Ca) and elements that behave
\textit{like} $\alpha$-elements in Galactic halo stars (Ti). Ti is an
average of \ion{Ti}{1} and \ion{Ti}{2}, except for Pal 1-I, which
includes only \ion{Ti}{1}---this average is not weighted by the number
of lines.  The total $\alpha$ abundance shown in Figure
\ref{fig:AlphaFe2} is an average of Mg, Ca, and Ti.  O was excluded
from this average because of its weak lines; Si was excluded because
our [Si/Fe] may be systematically too high (see Section
\ref{subsec:comparison}).  Note that the [$\alpha$/Fe] values from
\citet{Venn2004} and \citet{Pritzl} have been re-calculated using our
definition.

The \ion{O}{1} lines in our spectral range are the forbidden 6300 and
6363 \AA \hspace{0.025in} transitions.  Both of these lines are
extremely weak and had to measured with spectrum syntheses.  Only
upper limits are available for Pal 1-I, -II, and -III.  Still, the
single O abundance seems to follow the general trend of the other
$\alpha$-elements.  Pal 1-C's slightly low [O/Fe] value agrees with
the lower range of Galactic field star abundances and with the old
OCs, particularly Be 20 and 29.

The Mg abundances were determined from lines of varying strengths for
M67-141, Pal 1-III, and -C.  M67-141 agrees nicely with the field
stars.  Only one line was detectable in Pal 1-I and -II, and had to be
measured with spectrum syntheses.  This 5528 \AA \hspace{0.025in} line
is very strong in all the stars, though it is under the 200 m\AA
\hspace{0.025in} limit for the Pal 1 stars.  Given that the Pal 1-I
and -II abundances agree with -III and -C, they seem reliable.
Overall the [Mg/Fe] ratios in Pal 1 are lower than the Galactic field
stars, the bulge and disk GCs, and the old OCs, but in agreement with
the halo clusters Pal 12 and Ter 7.

The Si abundances were also determined from a wide range of lines, but
again only a single line (at 6155 \AA) was detectable in Pal 1-I and
-II.  This line is of intermediate strength in the other two Pal 1
stars, but again the Pal 1-I and -II abundances agree with the others.
As discussed in Section \ref{subsec:NLTE}, M67-141's [Si/Fe] ratio is
slightly higher than previous values in the literature; this suggests
that Pal 1's [Si/Fe] might also be slightly high.  Its current [Si/Fe]
value places it in agreement with the Galactic field stars, the halo
GCs, the bulge/disk cluster 47 Tuc, and the old OCs.

The Ca abundance was determined from four to eighteen lines for the
Pal 1 stars, with Pal 1-II having the lowest number.  All four of
these \ion{Ca}{1} lines are fairly strong ($>$ 100 m\AA) in Pal 1-II,
which has a slightly lower Ca abundance than the other three stars.
Again these [Ca/Fe] values agree with Galactic field stars, GCs, and
OCs, and the Sgr GCs.  M67-141, with 15 Ca lines (most of them strong)
matches the [Ca/Fe] values found in the field stars.

A wide assortment of \ion{Ti}{1} and\ion{}{2} lines were used,
of varying strengths.  Only three to four \ion{Ti}{1} lines were
detectable in the Pal 1-I and -II spectra; those lines were of strong
and intermediate strength.  The other spectra had considerably more
\ion{Ti}{1} lines. Because many of the \ion{Ti}{2} lines are located
in the blue regions of our spectra, which have lower SNR, no
\ion{Ti}{2} lines are detectable in Pal 1-I, and only two intermediate
strength lines are detectable in Pal 1-II.  Even with only two lines, Pal
1-II's \ion{Ti}{1} and\ion{}{2} abundances are in excellent
agreement, as are those of Pal 1-III (with 32 \ion{Ti}{2} lines), -C
(with 23), and M67-141 (with 33).  The Ti abundances for M67-141 again
follow the field star trend, though Pal 1's [Ti/Fe] ratios are
slightly lower than Galactic field stars, GCs, and OCs.  The Pal 12
and Ter 7 [Ti/Fe] ratios are also slightly higher than Pal 1.

Overall, M67-141's $\alpha$ abundance ratios agree well with the
Galactic field stars.  Pal 1 shows good agreement with the halo GCs
Pal 12 and Ter 7.  With the exception of Si and Ca, the [X/Fe] ratios
for the $\alpha$-elements in Pal 1 are slightly lower than the bulge
and disk clusters and the Galactic field stars.  The Galactic
[$\alpha$/Fe] ratios are distinct between the thick disk and the thin
disk stars; Pal 1 is clearly lower than the thick disk stars, but is
not clearly separated from the thin disk.  Thus, Pal 1's [$\alpha$/Fe]
ratios are not distinct from the Galactic field stars, GCs, and OCs
\textit{within the $1\delta$ errors};\footnote{Recall that $1\delta =
1\sigma/\sqrt{N_{\rm{lines}}}$ (see Section
\ref{subsubsec:statistical}).} however, we note that the Pal 1 stars
are generally lower than the average Galactic trend.

\subsection{Na and Al}\label{subsec:NaAl}
\begin{figure}
\epsscale{1.25}
\plotone{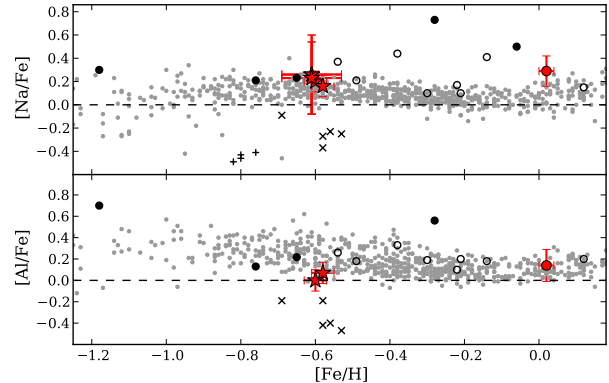}
\caption{[X/Fe] ratios of the light elements Na and Al versus
metallicity.  The points are as in Figure
\ref{fig:AlphaFe1}.\label{fig:NaAlFe}}
\end{figure}

Na and Al are also produced in massive stars during nucleosynthesis
through carbon burning and hydrogen shell burning
\citep{WoosleyWeaver}.  The [X/Fe] ratios are shown in Figure
\ref{fig:NaAlFe}.  Note that Galactic GCs all show star-to-star
variations in Na and Al, and the average abundances of the GCs may not
reflect the actual primordial abundances.  As discussed in Section
\ref{subsec:NLTE}, the Na lines with large NLTE corrections have been
avoided.  Thus, M67-141's Na abundance is based on the 6154/6161 \AA
\hspace{0.025in} doublet; the Pal 1 analysis also includes the
5682/5688 \AA \hspace{0.025in} doublet.  Both Pal 1 and M67-141 have
slightly higher [Na/Fe] values than most of the field stars, in
agreement with the old OCs and GCs.  The Pal 1 [Na/Fe] are distinctly
higher than in the halo GCs Pal 12 and Ter 7.

The Al abundances come from the 6696/6698 \AA \hspace{0.025in} lines,
both of which are weak in the Pal 1 spectra---thus, [Al/Fe] ratios are
not found for Pal 1-I and -II.  M67-141's [Al/Fe] agrees with the
field stars, while Pal 1's [Al/Fe] is slightly low for field stars,
bulge/disk GCs, and OCs. As with sodium above, the [Al/Fe] ratios in
the Pal 1 stars are distinct from those of stars in Ter 7.

Anticorrelations between Na/O and Al/Mg are seen in nearly all
Galactic GCs, and \citet{Carretta2010} suggest that all bona fide GCs
show a range in these elements.  In Pal 1 Al is clearly not enhanced,
and is slightly low for Galactic stars; without significant Al
overabundances Mg depletion is not expected to occur
\citep{Carretta2009}.  Na, however, \textit{is} slightly enhanced in
Pal 1, lying just above the Galactic field stars.  In the absence of
any stellar evolutionary effects, Na should behave like the
$\alpha$-elements, which are also distributed by Type II supernovae.
Thus Pal 1's high Na abundances may indicate that Na-Ne cycled gas is
present---since Al is not enhanced, this suggests that Al-Mg cycled
gas is not present.  With only upper limits on [O/Fe] for three stars,
we can only comment that Pal 1-C's [O/Fe] is slightly low, as expected
with higher [Na/Fe].  We further note that there is not a significant
range in the [Na/Fe] values for the four stars (see Section
\ref{subsubsec:properties}) which suggests that the high Na is
\textit{not} a sign of the canonical Na/O anticorrelation seen in
Galactic GCs---however, we cannot rule this out with only four stars.

\subsection{Iron-Peak Elements}\label{subsec:iron}
\begin{figure}
\epsscale{1.25}
\plotone{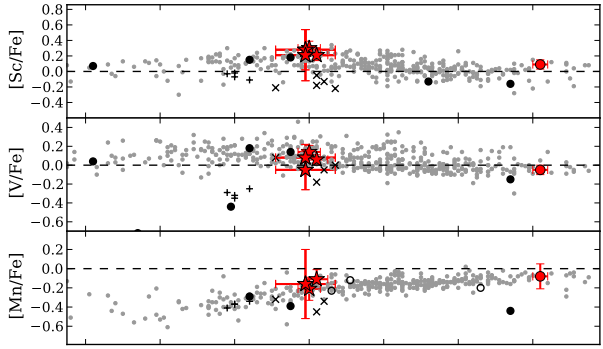}
\caption{[X/Fe] ratios of the iron-peak elements Sc, V, and Mn versus
metallicity.  The points are as in Figure
\ref{fig:AlphaFe1}.\\ \label{fig:ScVMnFe}}
\end{figure}

The iron-peak elements are formed in both Type Ia and Type II
supernovae, though the precise contributions of the different types
are unknown.  In Pal 1's metallicity regime Type Ia supernovae should
be the dominant contributors; \citet{Iwamoto1999} estimate that a Type
II supernova may create $\sim 0.1 \; \rm{M}_{\sun}$ of iron-peak
material while a Ia may contribute $\sim 0.8 \; \rm{M}_{\sun}$.
Figures \ref{fig:ScVMnFe} and \ref{fig:CoCrNiFe} show the [X/Fe]
ratios of the iron-peak elements. Note that in the single plot for
[Cr/Fe] both \ion{Cr}{1} (red) and \ion{Cr}{2} (yellow) are shown.

\vspace{0.4in}

\subsubsection{Scandium, Vanadium, Manganese, and Cobalt}\label{subsubsec:VCrMnCo}
The odd-Z elements are similar because they have only a single stable
isotope each; these elements also require corrections for hyperfine
splitting.  Note that Sc is included in this discussion, even though
it is not a traditional iron-peak element, as its production site is
primarily Type II supernovae.

\ion{Sc}{2} and \ion{V}{1} have various detectable lines in our
spectra.  Spectrum syntheses were performed on the 5527 \AA
\hspace{0.025in} \ion{Sc}{2} line for Pal 1-II and on the 6126 \AA
\hspace{0.025in} \ion{V}{1} line for -I.  Three intermediate strength
\ion{Sc}{2} lines (5667, 5669, and 6005 \AA) were detectable in Pal
1-I, while four intermediate strength \ion{V}{1} lines (6090, 6150,
6199, and 6225 \AA) were found in Pal 1-II.

Only four \ion{Mn}{1} lines were detectable and under the 200 m\AA
\hspace{0.025in} limit in M67-141, all of them strong ($>100$ m\AA).
As discussed in Section \ref{subsec:NLTE}, any NLTE corrections to
these lines should be small.  No \ion{Mn}{1} lines were measurable in
Pal 1-I, and only a single line (at 6022 \AA) was measured in Pal
1-II. The other two Pal 1 stars all have \ion{Mn}{1} lines with EWs
$<$ 80 m\AA, which should require small NLTE corrections.

\ion{Co}{1} has only two lines in this spectral range, at 5483 and
5647 \AA.  Neither line was detected in Pal 1-I or -II.  In M67-141,
Pal 1-III, and -C the 5483 \AA \hspace{0.025in} line is intermediate
or strong while the 5647 \AA \hspace{0.025in} line is weak to
intermediate.

All of Pal 1's odd iron-peak elements agree with the Galactic field
stars.  Sc is slightly higher than the general Galactic trend, putting
Pal 1 in closer agreement with 47 Tuc than with Ter 7 or Pal 12. The V
abundances are slightly lower than the Galactic trend, in agreement
with field stars and Ter 7.  [Mn/Fe] is slightly higher than all the
GCs, but agrees with field stars and the old OCs Be 20 and 29.
Finally, [Co/Fe] is slightly low, in agreement with the halo clusters
but still within range for OCs and field stars---a comparison cannot
be made with the GCs as Co abundances are not available for all
clusters.

\subsubsection{Chromium and Nickel}\label{subsubsec:Ni}
\begin{figure}
\epsscale{1.25}
\plotone{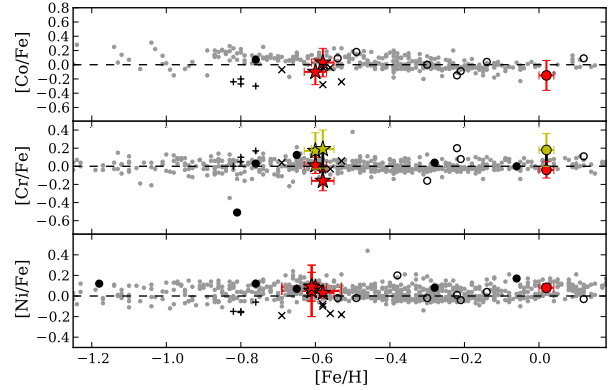}
\caption{[X/Fe] ratios of the iron-peak elements Co, Cr, and Ni versus
metallicity.  Both \ion{Cr}{1} and \ion{Cr}{2} are shown (\ion{Cr}{1}
in red, \ion{Cr}{2} in yellow); note that the low SNR stars Pal 1-I
and -II only have \ion{Cr}{1} abundances. The points are as in Figure
\ref{fig:AlphaFe1}.\\ \label{fig:CoCrNiFe}}
\end{figure}

There are 11 available \ion{Cr}{1} lines in our line list, but only
two \ion{Cr}{2} lines, both of which are in the blue spectral regions.
Thus, \ion{Cr}{2} abundances are not available for Pal 1-I and -II;
unfortunately, the redder \ion{Cr}{1} lines were also not measurable
in Pal 1-I and -II.  For the other stars the \ion{Cr}{1} abundances
are in good agreement with each other and with the literature.
M67-141 looks like Galactic field stars and GCs, while Pal 1 lies
slightly below the field stars and agrees with the halo GCs.  However,
the \ion{Cr}{2} abundances are considerably higher (identified as
yellow in Figure \ref{fig:CoCrNiFe}), an effect that has been observed
in previous studies of Galactic stars (e.g. \citealt{Zhang2009}).  The
\ion{Cr}{2} abundances may not be reliable because the lines are in
the blue regions of our spectra; however, the discrepancy between
\ion{Cr}{1} and \ion{Cr}{2} may be due to NLTE effects for \ion{Cr}{1}
\citep{Sobeck,BergemannCescutti}.  Therefore, neither [\ion{Cr}{1}/Fe]
nor [\ion{Cr}{2}/Fe] may be valuable choices for comparisons.

Ni also has many spectral lines, though only three and five are
detectable in Pal 1-I and -II, respectively.  Both M67-141 and Pal 1's
[Ni/Fe] values match those of field stars and clusters, with the
exceptions of Ter 7 and Pal 12, which are slightly lower.

\subsection{Copper and Zinc}\label{subsec:CuZn}
\begin{figure}
\epsscale{1.25}
\plotone{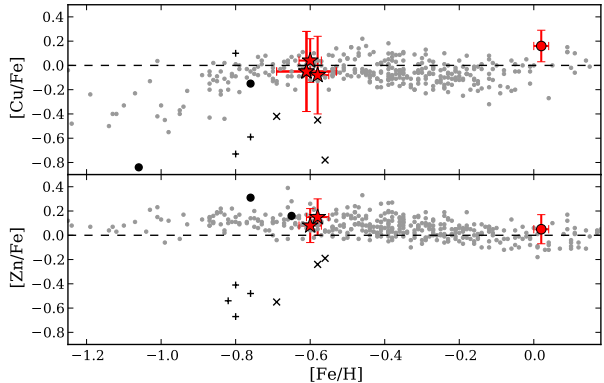}
\caption{[X/Fe] ratios of the elements Cu and Zn versus metallicity.
The points are as in Figure \ref{fig:AlphaFe1}.\\ \label{fig:CuZnFe}}
\end{figure}

The nucleosynthetic origins of Cu and Zn are varied, and the precise
yields from various sites depend on the models used.  Explosive
burning in Type Ia and II supernovae can create Cu and Zn, as can
neutron captures.  Given the uncertainty surrounding the formation of
these elements they are discussed separately from the rest.  Their
[X/Fe] ratios are shown in Figure \ref{fig:CuZnFe}.

Three \ion{Cu}{1} lines are examined here: 5105.5, 5700, and 5782 \AA.
These lines are all strong in M67-141, though the last two are $< 100$
\AA \hspace{0.025in} in the Pal 1 stars.  None are detectable in Pal
1-I, while only the last is detectable in Pal 1-II.  The 5782 \AA
\hspace{0.025in} line was just off the end of the red region in Pal
1-III and Pal 1-C, and the 5700 \AA \hspace{0.025in} line was obscured
by a cosmic ray in Pal 1-C.  The three Pal 1 [Cu/Fe] ratios agree with
the field stars.  However, the [Cu/Fe] values are \textit{not} within
$1\delta$ of Ter~7 and Pal~12.  M67-141's [Cu/Fe] is a bit high, but
its $1\delta$ errors place it with the Galactic field stars.

Three intermediate strength \ion{Zn}{1} lines lie in our spectral
region, at 4722, 4811, and 6362 \AA, and yield consistent abundances
when all three are measured in one star.  None are detectable in  Pal
1-I or -II, though all three are measurable in  M67-141, Pal 1-III,
and Pal 1-C. Similar to copper, both M67-141 and Pal 1 have [Zn/Fe]
ratios that agree with field stars and GCs but are higher than Ter~7
and Pal~12.

\subsection{r- and s-Process Elements}\label{subsec:rs}

The heavier elements are formed through neutron captures onto
iron-peak atoms, either via the rapid (r-) process in Type II
supernovae, or via the slow (s-) process, e.g. in low mass AGB stars.
In the Sun, the percentages of Y, Ba, La, Nd, and Eu that come from
the s-process are 75\%, 85\%, 72\%, 47\%, and 3\%, respectively
\citep{Burris2000}; thus Eu is an important r-process indicator.  The
[X/Fe] ratios for the elements that are primarily due to the s-process
(Y, Ba, and La) are shown in Figure \ref{fig:YBaLaFe}, while those
with larger r-process contributions (Nd and Eu) are shown in Figure
\ref{fig:NdEuFe}.

\subsubsection{s-Process Elements}\label{subsubsec:s}
Five \ion{Y}{2} lines are detectable in the Pal 1 spectra, all in the
blue: 4884, 4900, 5087, 5200, and 5403 \AA.  The first four lines are
of intermediate-strength in the Pal 1 stars while the last is
weak. All five lines yield consistent abundances when measured in one
star. Spectrum syntheses were performed on the 5087 and 5200 \AA
\hspace{0.025in} lines for Pal 1-I and on the 5200 and 5403 \AA
\hspace{0.025in} lines for -II; only upper limits are given.  While
M67-141 agrees well with the field stars, the Pal 1 stars are
deficient in [Y/Fe] compared to field stars and globular clusters. The
Pal 1 [Y/Fe] ratios are in better agreement with those from Ter~7 and
Pal~12.

\begin{figure}
\epsscale{1.25}
\plotone{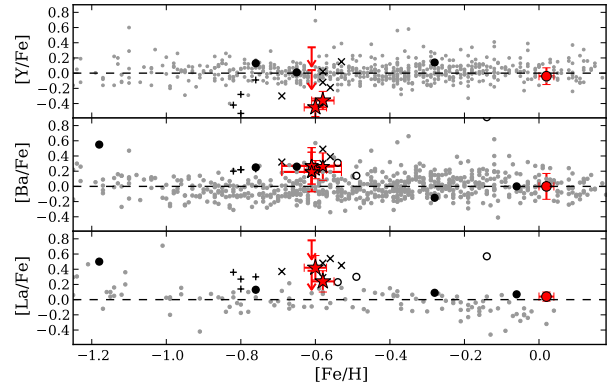}
\caption{[X/Fe] ratios of the elements Y, Ba, and La versus
metallicity.  The points are as in Figure
\ref{fig:AlphaFe1}, with additional Galactic abundances from
\citet{Simmerer}.\\ \label{fig:YBaLaFe}}
\end{figure}

\ion{Ba}{2} and \ion{La}{2} have three and four lines, respectively,
in our line list.  As mentioned in Section \ref{subsec:SS}, spectrum
syntheses were performed on all the 6142 \AA \hspace{0.025in} lines to
include the effect of the blended \ion{Ni}{1} line.  Another
\ion{Ba}{2} line, 4934 \AA, was quite strong and was not always below
the 200 m\AA \hspace{0.025in} limit.  The remaining \ion{Ba}{2} line,
5853 \AA, is of intermediate to strong strength.  The four \ion{La}{2}
lines (5302, 5304, 6320, and 6774 \AA) are all weak in the Pal 1
stars.  Spectrum syntheses were performed on the 6774 \AA
\hspace{0.025in} line to obtain upper limits for Pal 1-I and
-II. [Ba/Fe] and [La/Fe] show a similar trend in Figure
\ref{fig:YBaLaFe}: both are higher in Pal 1 than in the field stars,
but both are in agreement with 47~Tuc, Ter~7, Pal~12, and the old OCs,
particularly Be 29.

High s-process yields have been observed in Galactic GC (e.g. M4,
\citealt{Yong2008}), and have typically been explained by primordial
variations, i.e. the cluster happened to be born in a region where all
s-process yields were particularly high. However, in Pal 1 not all the
s-process elements are high: Y, a first-peak s-process element, shows
the opposite trend from Ba and La, which are second-peak elements.  To
compare the relative contributions of second-peak to first-peak,
[Ba/Y] vs. [Fe/H] is shown in the top panel of
Fig. \ref{fig:BaYBaEuEuAlphaFe}.  Though only lower limits are
available for Pal 1-I and -II, it is still evident that Pal 1 has
higher [Ba/Y] ratios than the field stars and GCs, though it is in
good agreement with the ratios for Ter~7 and Pal~12.

\subsubsection{r-Process Elements}\label{subsubsec:r}
Five \ion{Nd}{2} lines are observable in M67-141.  Of these five, only
two (5250 and 5320 \AA) are strong enough to be seen in the Pal 1
stars.  Only the latter line was detected in Pal 1-I; no \ion{Nd}{2}
lines were found in Pal 1-II.  The 6645 \AA \hspace{0.025in} \ion{Eu}{2}
line required spectrum syntheses for all stars; again, only upper
limits were obtained for Pal 1-I and -II.

\begin{figure}
\epsscale{1.25}
\plotone{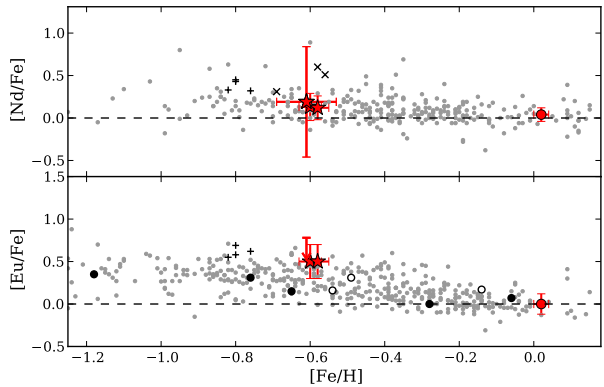}
\caption{[X/Fe] ratios of the elements Nd and Eu versus metallicity.
The points are as in Figure \ref{fig:YBaLaFe}.\label{fig:NdEuFe}}
\end{figure}

Figure \ref{fig:NdEuFe} shows that M67-141 follows the field star
trend for both Nd and Eu, while Pal 1, though slightly high, is still
in agreement with the Galactic field stars within its $1\delta$
errors.  Pal 1's [Nd/Fe] ratio may be a bit low compared to Ter~7
---no comparisons are available for the other Galactic GCs.  The
[Eu/Fe] of the Pal 1 stars is in good agreement with Pal~12, but those
are both much higher than in 47 Tuc and other Galactic GCs and OCs.

Since 97\% of Eu is produced via the r-process in the Sun
\citep{Burris2000}, it is primarily an r-process indicator. Also,
since the site of the r-process is believed to be Type II supernovae,
which also distribute the $\alpha$-elements, then the
$[\rm{Eu}/\alpha]$ ratio should be correlated.  The middle panel in
Figure \ref{fig:BaYBaEuEuAlphaFe} shows $[\rm{Eu}/\alpha]$
vs. [Fe/H]. Pal 1 has higher $[\rm{Eu}/\alpha]$ ratios than the
Galactic field stars, GCs, and OCs, but is in good agreement with the
$[\rm{Eu}/\alpha]$ ratios in Ter~7 and Pal~12. The excess Eu relative
to the Galactic stars suggests different yields for Eu and $\alpha$
elements, and that whatever the source of these different yields, it
was similar for Pal~1, Ter~7, and Pal~12.

\subsubsection{s-Process vs. r-Process}\label{subsubsec:svsr}
\begin{figure}
\epsscale{1.25}
\plotone{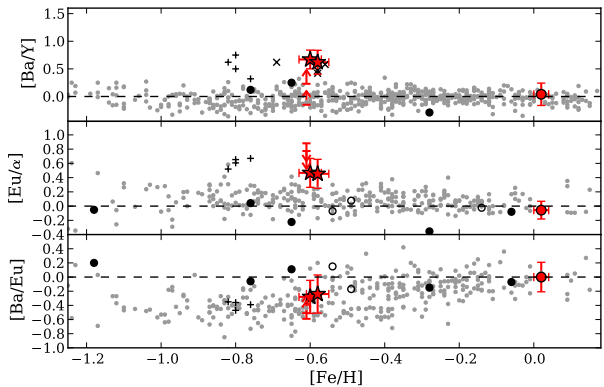}
\caption{[Ba/Y], [Ba/Eu], and [Eu/$\alpha$] versus metallicity.  The
points are as in Figure
\ref{fig:YBaLaFe}.\label{fig:BaYBaEuEuAlphaFe}}
\end{figure}

The ratio [Ba/Eu] provides a clue of the ratio of the s- to r-process
contributions.  The bottom panel in Figure \ref{fig:BaYBaEuEuAlphaFe}
shows that Pal~1 is in good agreement with the Galactic field star
distribution, suggesting that in Pal 1 the s-process contributes to
the chemical evolution in a similar way as in the Galaxy.  It is
interesting that Pal~1's [Ba/Eu] ratios are similar to Pal~12 and Be
20, but that those are all slightly lower than the rest of the
Galactic GCs and OCs.

\section{Discussion}\label{sec:Discussion}
With the derived abundances of twenty one different elements, the
guiding questions of this analysis can be addressed.  Ultimately we
wish to understand what type of cluster Pal 1 is and whether or not it
has been accreted from a satellite galaxy.  The answers to these
questions may have implications for galaxy and cluster formation.

\vspace{0.4in}
\subsection{What Kind of a Cluster is Pal 1?}\label{subsec:cluster}
As discussed in Section \ref{sec:Intro}, Pal 1 has traditionally been
classified as a GC, primarily because of its location above the
Galactic plane and its high concentration parameter.  However, Pal 1's
high [Fe/H] and red horizontal branch make it similar to the
bulge/disk GCs \citep{MackeyVandenbergh}, while its age sets it apart
from all GCs.  In fact, its age and [Fe/H] place it between the
classical definitions of GCs and OCs.  These unusual characteristics
have caused Pal 1 to be labeled as a ``transitional cluster''
\citep{Ortolani}.

In addition, Pal 1's brightness and size make it an ultra faint
cluster.  A graph of absolute magnitude versus half-light radius (see
Figure 1 in \citealt{NiedersteOstholt}) shows that Pal 1 is fainter
than typical globular clusters and occupies a position that could be
an extrapolation of the dwarf galaxy and ultra faint dwarf galaxy
trend. It is therefore worth investigating Pal 1's similarities to
these different types of objects.

\subsubsection{Basic Properties of Pal 1}\label{subsubsec:properties}
We begin by summarizing the basic properties of Pal 1 that have been
determined in this paper.  We have derived a average metallicity of
$[\rm{Fe/H}] = -0.60 \pm 0.01$ for Pal 1, a result that is in
agreement with both the CaT estimates \citep{RosenbergB} and the
isochrone fits \citep{Sarajedini}.  We further find that the cluster
is not $\alpha$-enhanced.  Defining $\alpha$ to be an average of Mg,
Ca, and Ti, the average $[\alpha/\rm{Fe}]\approx 0.0$.

A revised age can be derived using our updated, detailed abundances.
Isochrones from the Dartmouth Stellar Evolution Database \citep{DSED}
have been adopted to derive a new age for Pal 1 using the derived
$[\rm{Fe/H}] \approx -0.6$ and $[\alpha/\rm{Fe}]\approx 0.0$
(previously values of $[\rm{Fe/H}] = -0.7$ and $[\alpha/\rm{Fe}] =
0.2$ were used, \citealt{Sarajedini}). The same distance modulus and
reddening are used as for the photometric atmospheric parameters:
$(m-M)_0 = 15.65$ and $E(B-V) = 0.20$ (see Section
\ref{subsec:Spectroscopic} for details); the photometry is taken from
the \citet{Sarajedini} HST Globular Cluster
Treasury.\footnote{\url{http://www.astro.ufl.edu/$\sim$ata/public\_hstgc/databases.html}}

Fits of the \citet{DSED} isochrones are shown in Figure \ref{fig:CMD},
for ages of 4, 5, and 6 Gyr; the best age remains at $5\pm 1$ Gyr.
The differences between these fits and those of \citet{Sarajedini} are
very slight, as expected since the adopted [Fe/H] and [$\alpha$/Fe]
combinations correspond to roughly the same [M/H].  Thus, our analysis
confirms the previous findings that Pal 1 is indeed a young cluster.
We further note that none of our target red giant stars lie on the
isochrone RGBs; this may suggest that our targets are actually
horizontal branch stars.\footnote{If the stars \textit{are} horizontal
branch stars, we would expect to obtain slightly different surface
gravities, due to, e.g., mass loss.  However, such differences should
have a minor ($<0.1$ dex) effect on $\log g$.}

No conclusive signs of any abundance spreads are detected among the
four Pal 1 stars, suggesting that Pal 1 is indeed a simple stellar
population. Table \ref{table:spreads} shows the mean [X/Fe] ratios in
Pal 1 and the abundance spreads within the cluster, according to the
method outlined in \citet{Cohen}.  Here $\sigma$ is the dispersion
about the mean abundance in the cluster and $\sigma$(obs) is the mean
uncertainty of an element in a single star.  The spread ratio is then
a comparison of the width of the mean distribution to the width of a
single star's error estimate: large values of the spread ratio imply
that the star-to-star variations are larger than the average
uncertainty for an individual star, suggesting that the cluster has a
genuine abundance spread.  The Pal 1 stars do not show a significant
spread for most elements (a spread ratio $>1.0$ would be considered
significant; \citealt{Cohen}).  \ion{Cr}{1} has a spread ratio $>1.0$,
but abundances are only available for 2 stars.  The low spread
ratio for Na suggests that a Na/O anticorrelation is not present,
although O is available for only one star.  Thus, Pal 1 does not
appear to show any signs of star-to-star variations. This is further
confirmed by its CMD: multiple populations are not evident and a
single isochrone fits it well, implying that the cluster is coeval.
\begin{center}
\begin{deluxetable*}{lccccc}
\centering
\tabletypesize{\scriptsize}
\tablecolumns{6}
\tablewidth{0pc}
\tablecaption{Mean abundances and spread ratios for the four stars in
Pal 1, according to the method of \citet{Cohen}.  A spread ratio $>1$
indicates that a significant spread exists; Pal 1 shows no evidence
for any abundance spreads.  See the text for a description of how
these values are calculated.\label{table:spreads}}
\tablehead{
 & Mean [X/Fe]\tablenotemark{a} & Number of stars & $\sigma$ & $\sigma$(obs) & Spread Ratio}
\startdata
\ion{Fe}{1} & -0.60 & 4 & 0.01 & 0.06 & 0.26\\
\ion{Fe}{2} & -0.60 & 4 & 0.10 & 0.21 & 0.46\\
Na & 0.21 & 4 & 0.04 & 0.21 & 0.21\\
Mg & -0.06 & 4 & 0.07 & 0.16 & 0.44\\
Al & 0.04 & 2 & 0.05 & 0.10 & 0.49\\
Si & 0.16 & 4 & 0.07 & 0.15 & 0.45\\
Ca & 0.10 & 4 & 0.09 & 0.13 & 0.75\\
Sc & 0.25 & 4 & 0.05 & 0.19 & 0.25\\
\ion{Ti}{1} & -0.06 & 4 & 0.07 & 0.21 & 0.33\\
\ion{Ti}{2} & -0.04 & 3 & 0.11 & 0.21 & 0.52\\
V & 0.06 & 4 & 0.08 & 0.12 & 0.68\\
\ion{Cr}{1} & -0.09 & 2 & 0.12 & 0.10 & 1.20 \\
\ion{Cr}{2} & 0.18 & 2 & 0.01 & 0.21 & 0.07\\
Mn & -0.16 & 3 & 0.05 & 0.18 & 0.27\\
Co & -0.04 & 2 & 0.09 & 0.20 & 0.47\\
Ni & 0.06 & 4 & 0.03 & 0.12 & 0.22\\
Cu & -0.03 & 3 & 0.06 & 0.28 & 0.23\\
Zn & 0.12 & 2 & 0.05 & 0.15 & 0.34\\
Y & -0.41 & 2 & 0.06 & 0.12 & 0.53\\
Ba & 0.24 & 4 & 0.04 & 0.20 & 0.18\\
La & 0.33 & 2 & 0.13 & 0.15 & 0.85\\
Nd & 0.15 & 3 & 0.04 & 0.32 & 0.12 \\
Eu & 0.50 & 2 & 0.0 & 0.20 & 0.0
\enddata
\tablenotetext{a}{[X/H] is given instead for \ion{Fe}{1} and
\ion{Fe}{2}.}
\end{deluxetable*}
\end{center} 

\vspace{-0.4in}
Thus, our analysis finds Pal 1 to be a metal-rich,
non-$\alpha$-enhanced, young, chemically homogeneous stellar
population.

\subsubsection{Globular or Open Cluster?}\label{subsubsec:GCvsOC}
Given that the cluster types are not well-defined it is difficult to
distinguish between the two types of clusters, particularly when
examining chemical abundances.  However, multiple populations have
been observed in nearly all Galactic GCs, but not in open clusters.
Thus, signs of the Na/O and/or the Mg/Al anticorrelations would be
positive indicators for a GC.

No signs of the Mg/Al anticorrelation are detected in Pal 1, and O
abundances are only available for one of the four stars, Pal 1-C. As
discussed in Section \ref{subsec:NaAl}, while Na is slightly high and
O is slightly low in Pal 1-C, there is no significant range in the Na
abundances, and the four stars do not show evidence for an
anticorrelation. Certain models (e.g. \citealt{ConroySpergel}) suggest
that the formation of a second population in a GC depends on the
cluster mass, since more massive clusters can retain more gas to form
a second population---Pal 1's current mass, $\log(M/\rm{M}_{\sun})
\approx 3.2$ \citep{NiedersteOstholt}, is below the critical mass for
retaining the gas to form a second population. It has been known for
some time that Pal 1 is in the process of evaporating
\citep{RosenbergA}; \citet{NiedersteOstholt} further show that the
mass in Pal 1's tidal tails is roughly equal to its current mass,
suggesting that the cluster might have been at least twice as massive
in the past.  In the \citet{ConroySpergel} context, however, Pal 1's
lack of a second generation implies that it was \textit{never} more
massive than $\log(M/\rm{M}_{\sun}) = 4.0$, regardless of its current
size.  This restriction is not necessarily unusual for a GC: the GC
Pal 12 does not show signs of abundance spreads \citep{Cohen} and five
intermediate age ($<2$ Gyr) LMC GCs show no evidence for multiple
populations in their CMDs.

Pal 1's detailed abundances can also be compared to the
well-established GCs in the Galaxy.  Its metallicity alone puts it in
an unusual regime: while there are many metal-rich GCs, those are
usually centrally concentrated in the bulge or disk, while Pal 1 is
located much further from the center and far above the disk
($R_{\rm{GC}} = 17.2$ kpc and $Z=3.6$ kpc;
\citealt{Harris}, 2010 edition).\footnote{Recall that previous
classifications have labeled Pal 1 as a bulge/disk cluster
(e.g. \citealt{MackeyVandenbergh}) based on its [Fe/H] and HB
morphology rather than its location, but its unusual location does
distinguish it from the bulge/disk GCs.}  Most of the iron-group
(Figures \ref{fig:ScVMnFe} and \ref{fig:CoCrNiFe}), $\alpha$- (Figures
\ref{fig:AlphaFe1} and \ref{fig:AlphaFe2}), and neutron capture
elements (Figures \ref{fig:YBaLaFe}, \ref{fig:NdEuFe}, and
\ref{fig:BaYBaEuEuAlphaFe}) are in good agreement between Pal~1 and
the Sgr clusters; however, the Na, Al, Cu, and Zn abundances (Figures
\ref{fig:NaAlFe} and \ref{fig:CuZnFe}) are distinctly different. Thus,
the chemical pattern of Pal 1 is not clearly similar to either the
metal-rich Galactic GCs or the Sgr GCs. Of the GCs, Pal~1 is most
similar to Pal~12 and Ter~7, both known to be associated with the Sgr
dSph.

There are several old, metal-poor OCs with detailed abundances in the
literature, though most are more metal-rich than Pal 1.  Several of
these clusters are located in the outer disk, and may have been
accreted from a dwarf galaxy or formed during a major merging event
\citep{Yong}.  Be 20 and 29, the two clusters in our sample that lie
closest to Pal 1's metallicity, have similar abundance ratios as the
Galactic stars, with the exceptions of the s-process neutron capture
elements. Thus, the only elements that are clearly discrepant between
Be 20/29 and Pal 1 are the $\alpha$-elements (Figures
\ref{fig:AlphaFe1} and \ref{fig:AlphaFe2}) and Eu (and by extension
[Eu/$\alpha$] and [Ba/Eu]; Figures \ref{fig:NdEuFe} and
\ref{fig:BaYBaEuEuAlphaFe}).  Cu, Zn, and Y abundances are not
available for Be 20 and 29; however, Pal 1's Na and Al abundances
(Figure \ref{fig:NaAlFe}) agree better with Be 20 and 29 than with the
Sgr clusters.

It is then natural to ask if Pal 1 looks more like the GCs Pal 12 and
Ter 7 or the OCs Be 20 and 29.  In terms of the $\alpha$-elements and
neutron capture elements Pal 1 looks more like the GCs, while Na and
Al agree more with the OCs.  However, \textit{chemically} comparing
Pal 1 to the two cluster types is not an accurate way of determining
Pal 1's cluster type, as Pal 1, Be 20, and Be 29 are not conclusively
associated with dwarf galaxies while Pal 12 and Ter 7 are associated
with the Sgr dSph.  Furthermore, the chemical abundances of stars
depend on their host galaxies' star formation histories, etc., and
these quantities are unique to individual galaxies.  Thus, unless all
of the clusters originated in the Sgr dSph, this comparison is not
straightforward.  We leave the question of Pal 1's origin to Section
\ref{subsec:extragalactic} and we therefore focus on other parameters
that may distinguish GCs from OCs.

Table \ref{table:other} shows various quantities that can be compared
between the clusters: the Galactocentric radius, $R_{\rm{GC}}$ (in
kpc); the distance above the Galactic plane, $Z$ (in kpc); distance
from the Sun, $d$ (in kpc); the age (in Gyr; though note that these
ages are assembled from different sources, and may not be valid for
comparisons); metallicity, [Fe/H]; absolute visual magnitude, $M_{V}$;
half-light radii, $r_{h}$ (in pc; note that the distance was used to
convert the half-light radius from arcminutes to pc); and
concentration parameter, $c$.  The last two parameters are only
available for the GCs.  Note that we also include Whiting 1, a GC
associated with Sgr \citep{Carraro}, though there are no chemical
abundances available for this cluster at this time. The values of
$R_{\rm{GC}}$ are similar for all the clusters in the table.  The OCs,
while more than 1 kpc away from the plane, are still closer than the
GCs, with Pal 1 lying between the two.  The ages and metallicities are
similar, but again Pal 1 is still quite young for a GC.  Though Pal 1
is fainter than the OC Be 29, its brightness is comparable to the GC
Whiting 1.  Finally, Pal 1 has a very small half-light radius compared
to Pal 12 and Ter 7, but in agreement with Whiting 1, and Pal 1 is
very concentrated, much more so than Ter 7 and Whiting 1.

\begin{center}
\begin{deluxetable*}{lccccccccccc}
\centering
\tabletypesize{\scriptsize}
\tablecolumns{12}
\tablewidth{0pc}
\tablecaption{Various parameters for the GCs and OCs used in this
study.  If no references are given then the data comes from this study
(in the case of Pal 1) or the \citet{Harris} catalog (2010 edition).
Note that the ages are assembled from different sources, and therefore
might not be directly comparable.  All values of $|Z_{\rm{Sgr}}|$ are
from \citet{LawMajewski}.\label{table:other}}
\tablehead{
Cluster & Traditional & $R_{GC}$ & $Z$ & $d$ & age & [Fe/H] &
$M_{V}$ & $r_{h}$\tablenotemark{a} & $c$ &
$|Z_{\rm{Sgr}}|$ & References\tablenotemark{b} \\
 & Classification & (kpc) & (kpc) & (kpc) & (Gyr) & & & (pc) & & (kpc)
& }
\startdata
Pal 1 & ? & 17.2 & 3.6 & 14.2 & 5.0 & -0.6 & -2.52 & 1.8 & 2.57 & 9.8 &  \\
Pal 12 & GC & 15.8 & -14.1 & 19.0 & 9.5 & -0.80 & -4.47 & 9.1 & 2.99 & 3.24 & 1, 2\\
Ter 7 & GC & 15.6 & -7.8 & 22.8 & 8.0 & -0.60 & -5.01 & 4.9 & 0.93 & 1.09 & 1, 3\\
Whiting 1 & GC & 34.5 & -26.3 & 30.1 & 6.5 & -0.65 & -2.46 & 1.8 & 0.55 & 0.215 & 4 \\
Be 20 & OC & 15.8 & -2.5 & - & 5.8 & -0.44 & -2.06 & - & - & - & 5, 6, 7\\
Be 29 & OC & 21.6 & 1.8 & - & 4.5 & -0.54 & -4.64 & - & - & 1.79 & 4, 6, 8\\
Saurer 1 & OC & 19.2 & 1.7 & - & 5.0 & -0.38 & - & - & - & 5.25 & 4, 9\\
\enddata
\tablenotetext{a}{The distance $d$ was used to convert the half-light
radius ($r_h$) from arcminutes to parces.}
\tablenotetext{b}{\textbf{References.} (1) \citet{Dotter2010}; (2)
\citet{Cohen}; (3) \citet{Sbordone}; (4) from the sources assembled by
\citet{ForbesBridges}; (5) \citet{Andreuzzi}; (6) \citet{Yong}; (7)
\citet{DeMarchi2006}; (8) \citet{Lata2002}; (9) \citet{Frinchaboy2006}\\}
\end{deluxetable*} 
\end{center}
\vspace{-0.2in}

The distinctions  between OC and GC in Table \ref{table:other} seem to
be based primarily on distance from the Galactic plane.  Considering
that Pal 12, Ter 7, and Whiting 1 have been accreted by Sgr, this is
not a particularly valid criterion.  Consequently, the distance above
the Sgr plane, $|Z_{\rm{Sgr}}|$, is also plotted for clusters that are
(or might be) associated with the Sgr dSph (from
\citealt{LawMajewski}; note that Pal 1's $|Z_{\rm{Sgr}}|$ is also
included, although it is most likely \textit{not} associated with
Sgr).  Examined this way, the clusters look very similar,
i.e. \textit{the young, metal-rich GCs that are close to the Sgr plane
are similar to the old, metal-poor OCs that are close to the Galactic
plane, suggesting that these clusters may be classified as open (or
intermediate-aged) clusters in the Sgr frame of reference.}
\citet{Carretta2010} also cast doubt upon the GC classification for
Pal 12 and Ter 7, given the lack of a definite Na/O anticorrelation.
Under the Carretta et al. definition we are forced to conclude that
Pal 1 is not a bona fide GC; however, it is not an obvious Galactic OC
either,  and therefore remains as an unusual cluster.

\subsubsection{A Connection with Ultra-Faint Dwarfs?}\label{subsubsec:UFD}
Given Pal 1's identification as an ultra faint cluster and its
proximity to the ultra-faint dwarfs (UFDs) in a plot of absolute
magnitude vs. half-light radius (Figure 1 in
\citealt{NiedersteOstholt}) it is tempting to consider  whether Pal 1
is related to these other ultra-faint objects.  The UFDs are
distinguished by their larger-than-expected velocity dispersions,
which imply high mass/light ratios and large amounts of dark matter.
The velocities for the five Pal 1 stars (including Pal 1-IV) listed in
Table \ref{table:observations} imply a velocity dispersion of $3.6\pm
1.5$ km s$^{-1}$ (using the formula from \citealt{Walker2006}),
\textit{assuming that all five velocities are orbital velocities of
cluster members.} Under this assumption, Pal 1's velocity dispersion
is in agreement with the velocity dispersions of several UFDs,
including Segue II, Leo V, Leo IV, and Hercules (see the summary of
UFD properties by \citealt{McConnachieCote}).

As tempting as this comparison may be, it is not altogether
appropriate.  Firstly, we now know that Pal 1 has tidal tails and is
therefore not in dynamical equilibrium \citep{NiedersteOstholt}.
Secondly, only one star (Pal 1-IV) has a discrepant velocity, and this
difference is easily explained if Pal 1-IV is a
binary; \citet{McConnachieCote} showed that the presence of binaries
can significantly boost the velocity dispersion of a low-mass system.
Alternatively, Pal 1-IV could be a nonmember---its location in the CMD
(Figure \ref{fig:CMD}) is separate from the other stars---however its
consistent chemistry and very close radial velocity make this seem
unlikely.  Therefore, Pal 1 remains an ultra faint cluster, and though
it would be interesting if it were an extrapolation of the UFDs, its
current velocity dispersion is best explained by its tidal disruption
or by the presence of a binary.

\subsection{Is Pal 1 Extragalactic?}\label{subsec:extragalactic}

Pal 1 has been tentatively associated with both the Galactic Anticenter
Stellar Stream (GASS) \citep{Crane} and the Canis Major (CMa)
overdensity (e.g. \citealt{Martin}, \citealt{ForbesBridges}), two
streams of stars that could be from disrupted dwarf galaxies.  If it
is associated with either of these systems, then Pal 1 should have
similar chemical abundances to their associated field stars and
clusters and to the stars in other low mass systems.

\begin{figure}
\epsscale{1.25}
\plotone{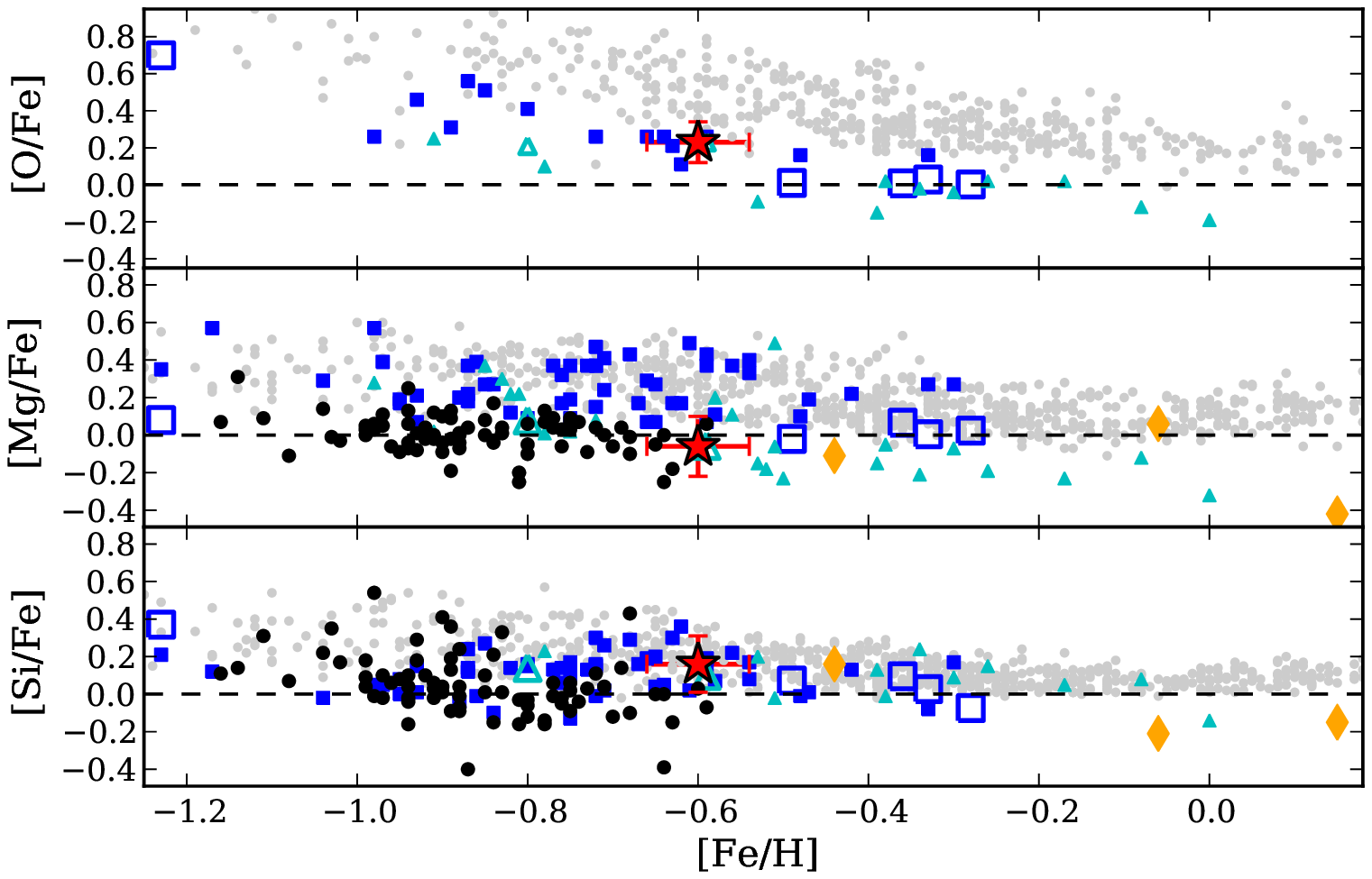}
\caption{[X/Fe] ratios of $\alpha$-elements O, Mg, and Si versus
metallicity for the stars in our study as well as Galactic field stars
(grey; as in Figure \ref{fig:AlphaFe1}) and stars in other galaxies.
Black points are Fornax stars \citep{Letarte2010}; cyan triangles are
Sagittarius (Sgr) stars (field stars are from \citealt{Sbordone2007},
\citealt{Monaco}, and \citealt{ChouSgr}, while GC stars are from
\citealt{Cohen} and \citealt{Sbordone}); blue squares are Large
Magellanic Cloud stars (field stars from \citealt{Pompeia}; cluster
stars from \citealt{Johnson2006} and \citealt{Mucciarelli2008}); and
orange diamonds are field stars in the Canis Major overdensity
\citep{SbordoneCMa}.\label{fig:AlphaFe1DG}}
\end{figure}

\begin{figure}
\epsscale{1.25}
\plotone{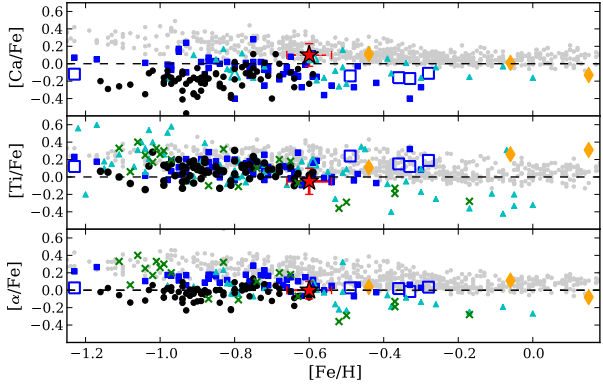}
\caption{[X/Fe] ratios of the $\alpha$-elements Ca and Ti, as
well as of the average $\alpha$ from Mg, Ca, and Ti.  The Ti abundance
is an average of \ion{Ti}{1} and \ion{Ti}{2}, except for Pal 1-I,
whose \ion{Ti}{1} abundance is shown since its \ion{Ti}{2} abundance
is abnormally high.  The points are as in Figure
\ref{fig:AlphaFe1DG}, with the addition of Galactic Anticenter Stellar
Stream (GASS; green crosses) stars from
\citet{Chou}.\label{fig:AlphaFe2DG}}
\end{figure}

\begin{figure}
\epsscale{1.25}
\plotone{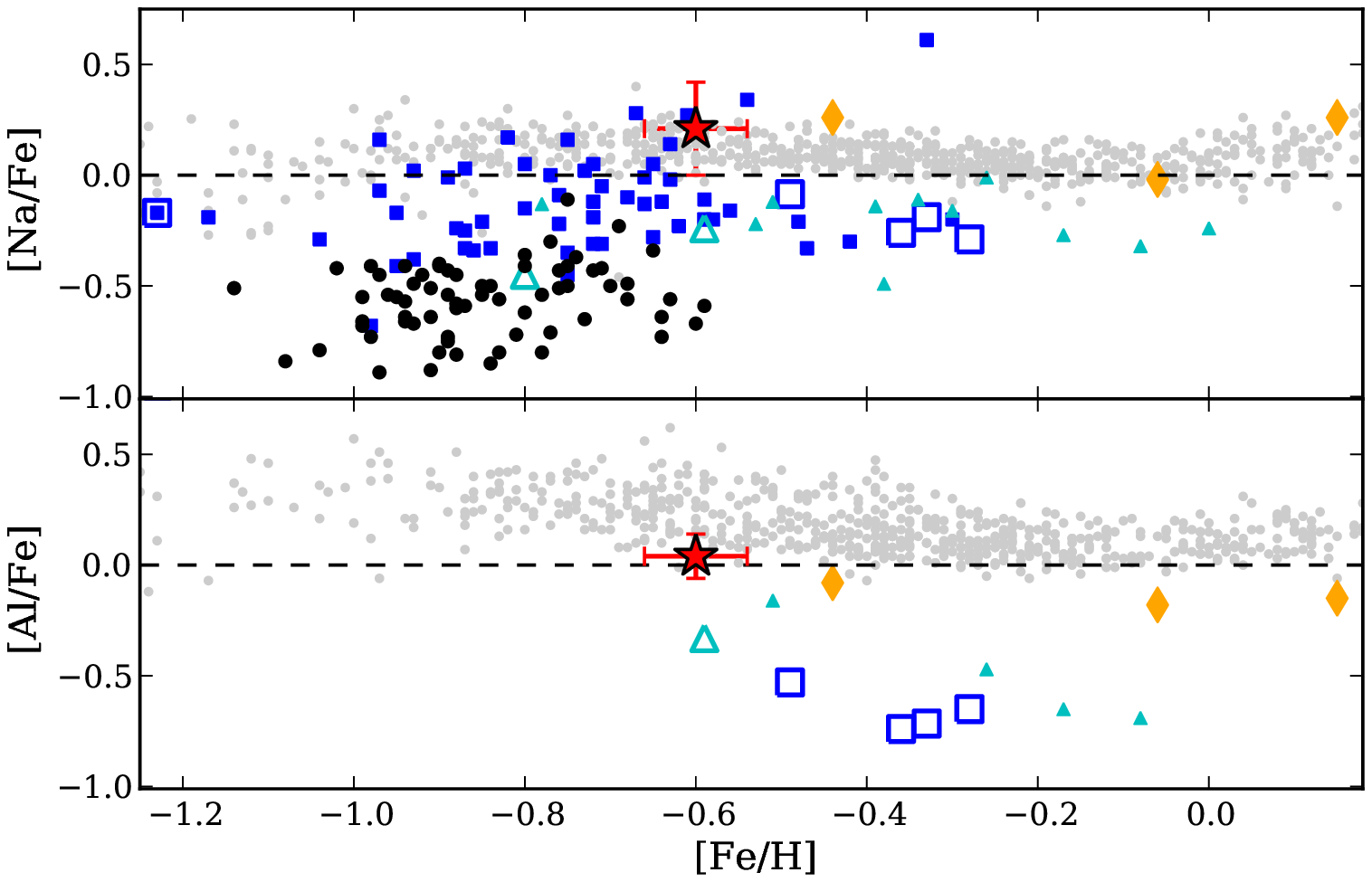}
\caption{[X/Fe] ratios of the light elements Na and Al versus
metallicity.  The points are as in Figure
\ref{fig:AlphaFe1DG}.\label{fig:NaAlFeDG}}
\end{figure}

\begin{figure}
\epsscale{1.25}
\plotone{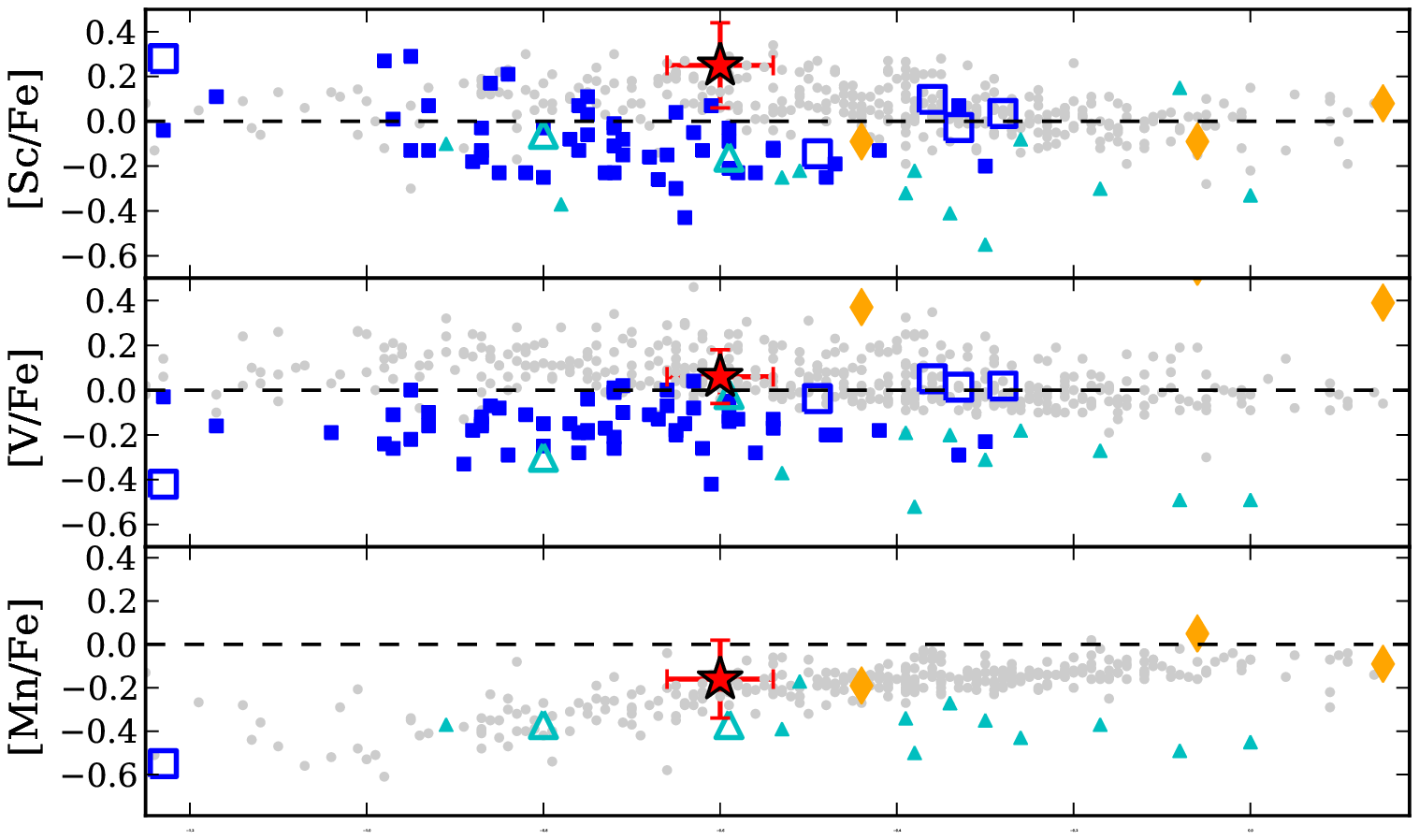}
\caption{[X/Fe] ratios of the iron-peak elements Sc, V, and Mn versus
metallicity.  The points are as in Figure
\ref{fig:AlphaFe1DG}.\label{fig:ScVMnFeDG}}
\end{figure}

\begin{figure}
\epsscale{1.25}
\plotone{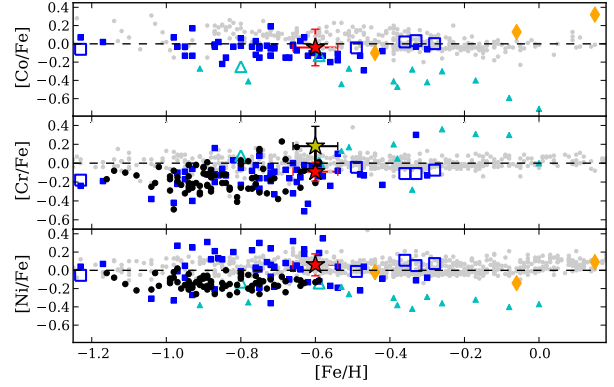}
\caption{[X/Fe] ratios of the iron-peak elements Co, Cr, and Ni
versus metallicity.  Both \ion{Cr}{1} and \ion{Cr}{2} are shown
(\ion{Cr}{1} in red, \ion{Cr}{2} in yellow).  The points are as in
Figure \ref{fig:AlphaFe1DG}.\label{fig:CoCrNiFeDG}}
\end{figure}

\begin{figure}
\epsscale{1.25}
\plotone{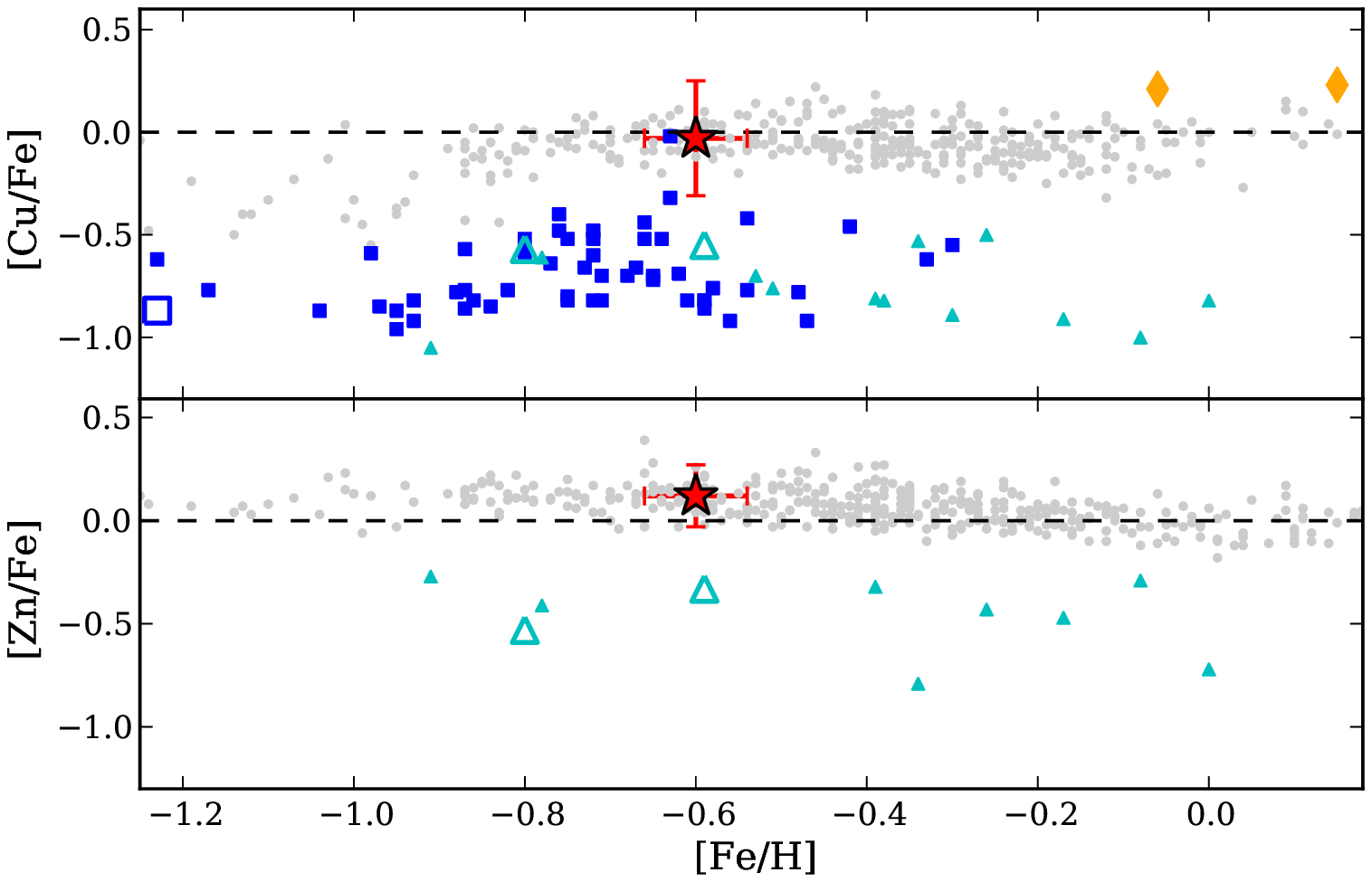}
\caption{[X/Fe] ratios of the elements Cu and Zn versus
metallicity.  The points are as in Figure 
\ref{fig:AlphaFe1DG}.\label{fig:CuZnFeDG}}
\end{figure}

\begin{figure}
\epsscale{1.25}
\plotone{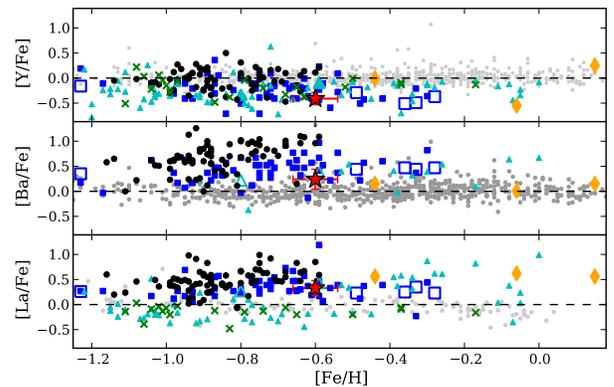}
\caption{[X/Fe] ratios of the elements Y, Ba, and La versus
metallicity.  The points are as in Figure
\ref{fig:AlphaFe2DG}.\label{fig:YBaLaFeDG}}
\end{figure}

\begin{figure}
\epsscale{1.25}
\plotone{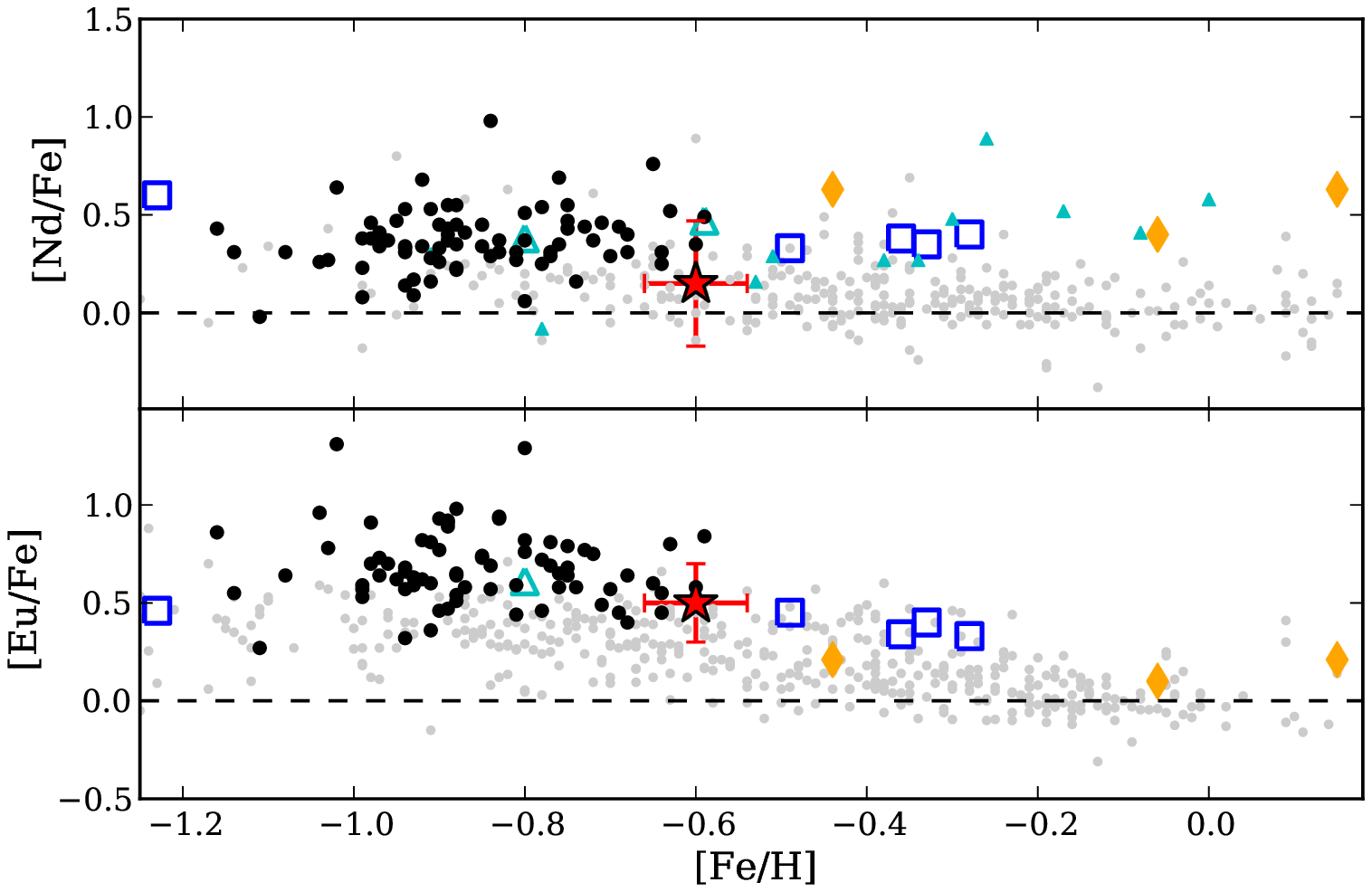}
\caption{[X/Fe] ratios of the elements Nd and Eu versus metallicity.
The points are as in Figure \ref{fig:AlphaFe1DG}.\label{fig:NdEuFeDG}}
\end{figure}

\begin{figure}
\epsscale{1.25}
\plotone{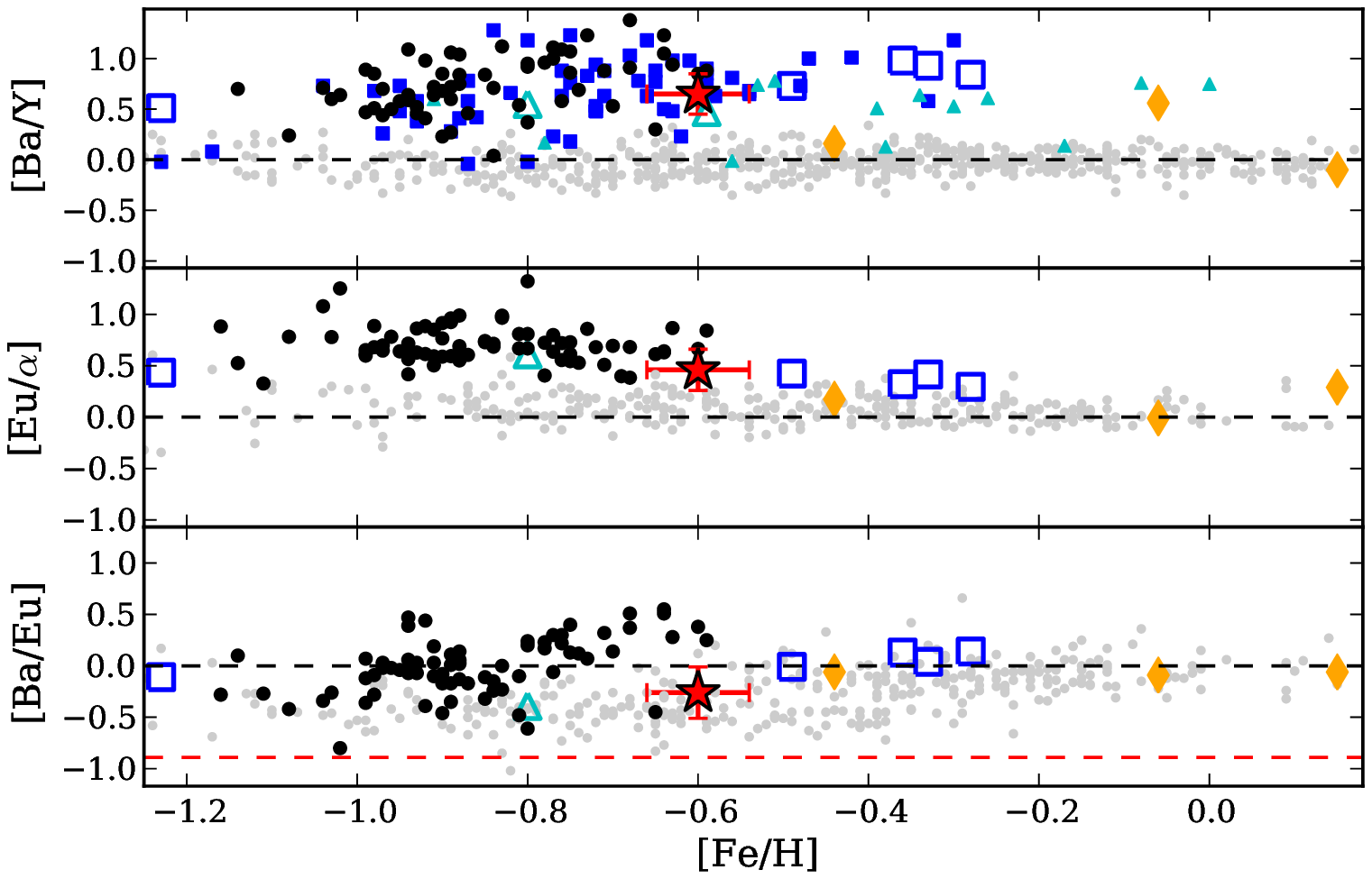}
\caption{[Ba/Y], [Ba/Eu], and [Eu/$\alpha$] versus metallicity.  The
dashed red line shows the r-process-only [Ba/Eu] ratio (from
\citealt{Burris2000}).  The points are as in Figure
\ref{fig:AlphaFe1DG}.\label{fig:BaYBaEuEuAlphaFeDG}}
\end{figure}

Figures \ref{fig:AlphaFe1DG} through \ref{fig:BaYBaEuEuAlphaFeDG}
compare the Pal 1 abundances to those of stars in the nearby
dwarf galaxies.  Note that only the average Pal 1 abundance (red star
symbol) is plotted, with the mean abundance error for [X/Fe]
(e.g. $\sigma$(obs) in Table \ref{table:spreads}). The Galactic field
stars are shown in grey.  Dwarf galaxies are also included, with their
field stars as small symbols and their GCs as larger open symbols.
Again, abundances have been averaged over an entire cluster.  Black
points are Fornax stars \citep{Letarte2010}; cyan triangles are
Sagittarius (Sgr) stars (field stars are from \citealt{Sbordone2007},
\citealt{Monaco}, and \citealt{ChouSgr}, while GC stars are from
\citealt{Cohen} and \citealt{Sbordone}); blue squares are Large
Magellanic Cloud stars (field stars from \citealt{Pompeia}; cluster
stars from \citealt{Johnson2006} and \citealt{Mucciarelli2008}); green
crosses are Galactic Anticenter Stellar Stream (GASS; \citealt{Chou})
stars; and orange diamonds are field stars in the Canis Major
overdensity \citep{SbordoneCMa}.  While other dwarf galaxy data are
available in this metallicity range, plotting all of it makes these
plots nearly illegible, and we have therefore included only the most
relevant dwarf data (i.e. galaxies with many stars and many elemental
abundances in the same metallicity range as Pal 1) for this
discussion.

\subsubsection{General Comparisons: Dwarf Galaxy or MW cluster?}\label{subsubsec:general}

In terms of the $\alpha$-elements (Figures \ref{fig:AlphaFe1DG} and
\ref{fig:AlphaFe2DG}), Pal 1's slightly low [$\alpha$/Fe] ratios place it
with the dwarf galaxies (with the possible exception of Ca). This
suggests that Type Ia supernovae began to contribute to Pal 1's host
environment at a lower [Fe/H] than for stars in the Galaxy.  In
general, Pal 1's [$\alpha$/Fe] ratios agree with the Sgr and LMC
clusters.  The slightly subsolar [Ti/Fe] value agrees with the GASS
stars while the slightly low [Mg/Fe] and [Ti/Fe] and slightly higher
[Ca/Fe] and [Si/Fe] ratios agree with the $[\rm{Fe/H}]=-0.44$ CMa
star.  The latter CMa star is also similar to Pal 1 in Na and Al
(Figure \ref{fig:NaAlFeDG}); for the few available [Al/Fe] abundances
Pal 1 is higher than the Sgr and LMC clusters.

Of the iron-peak elements (Figures \ref{fig:ScVMnFeDG} and
\ref{fig:CoCrNiFeDG}), Pal 1's [Sc/Fe] remains clearly distinct from
the dwarf galaxy stars.  Pal 1's [V/Fe] ratio is in good agreement
with the dwarfs, with the exception of the CMa stars, which may be due
to HFS corrections.  Overall the Pal 1 Fe-peak elements show the best
agreement with the Galactic stars rather than the dwarf galaxy stars.
Variations in [Fe-peak/Fe] with metallicity have been suggested to be
due to metallicity-dependent supernovae yields, e.g. metal-poor Type
Ia supernovae will produce less Mn than metal-rich ones
\citep{Cescutti2008}, as seen in the Sgr field stars.  However, Pal
1's agreement with the Galactic and dwarf galaxy stars at a similar
metallicity suggests no significant dispersion in the Type Ia
contributions.  The more peculiar elements, Cu and Zn (Figure
\ref{fig:CuZnFeDG}), are in better agreement with the Galactic stars
and possibly CMa, and are clearly distinct from the Sgr and LMC
stars.

Finally, Pal 1's neutron capture elements (Figures
\ref{fig:YBaLaFeDG}, \ref{fig:NdEuFeDG}, and
\ref{fig:BaYBaEuEuAlphaFeDG}) agree better with the dwarf galaxies
than with the Galactic stars.  Pal 1's low [Y/Fe] and high [Ba/Fe],
[La/Fe], and [Ba/Y] ratios suggest that like the Sgr dSph, Pal 1 was
enriched by metal-poor AGB stars.  (This effect is also seen in the
[La/Y] ratios of the Sgr stars; \citealt{McWSmH}).  With an excess of
neutrons per Fe atom in metal-poor AGB stars, more heavy second-peak
s-process elements such as Ba and La can be created, leaving a deficit
of first-peak elements like Y \citep{Gallino,Bisterzo}.  This further
suggests that Pal 1 has not been enriched by its own metal-rich AGB
stars, which is perhaps not surprising given its lack of evolved stars
(see Figure 1). The GASS field stars do not have high [La/Fe] like Pal
1, suggesting that it does not have an excess of second-peak to
first-peak s-process elements.  The CMa stars do seem to have high
[Ba/Fe] and [La/Fe], and arguably have low [Y/Fe], in agreement with
Pal 1.

Pal 1 also has slightly higher [Eu/Fe], [Nd/Fe], and [Eu/$\alpha$]
ratios than Galactic stars, in agreement with the LMC, Sgr, and
Fornax.  While the CMa stars do show an excess of Nd, they do not
appear to have an excess of Eu, and CMa therefore has a normal value
for [Eu/$\alpha$].  This suggests that dwarf galaxies have an
additional source of r-process elements compared to the Galaxy (see
e.g. \citealt{Letarte2010}).  This is further supported by Pal 1 and
the dwarf galaxies' high [Eu/$\alpha$] ratio; the standard model
suggests that massive Type II supernovae are responsible for
dispersing the $\alpha$-elements, while lower mass (8-10 M$_{\sun}$)
Type II supernovae are the sole site of r-process elements like Eu.
If the dwarf galaxies had similar conditions to the Galaxy, then the
[Eu/$\alpha$] ratio should remain flat.

Despite its high values for [Ba/Fe] and [Eu/Fe], the [Ba/Eu] ratio in
Pal 1 agrees better with the Galactic stars than with the dwarf
galaxies, though the Fornax dispersion and Pal 12 do extend down to
Pal 1's value.  Again CMa's [Ba/Eu] values are in fair agreement with
Pal 1.

As discussed earlier, at solar metallicity Ba is primarily an
s-process element, while Eu is primarily an r-process element
\citep{Burris2000}.  The ratio [Ba/Eu] should therefore provide an
indication of the relative contributions of the s-process to the
r-process.  Before the s-process begins to contribute Ba is initially
produced solely through the r-process; as time goes on the s-process
creates more Ba but little Eu, and [Ba/Eu] increases.  The theoretical
r-process--only lower limit for [Ba/Eu] is shown in Figure
\ref{fig:BaYBaEuEuAlphaFeDG}.  Pal 1's slightly high Ba and Eu
abundances lead to a [Ba/Eu] value that agrees with the classical
prediction for a system that has had little contribution from the
s-process.  Clearly this cannot be the case: Pal 1's [$\alpha$/Fe]
ratios show that Type Ia supernovae have contributed to the system,
and therefore AGB stars must have also contributed since the timescale
for evolution of Type Ia supernovae and AGB stars is thought to be
similar in chemical evolution models (e.g. \citealt{Travaglio2004},
\citealt{Matteucci}, \citealt{Zolotov}).  The [Ba/Y] ratio further
implies that metal-poor AGB stars must have contributed to the system.
Thus, the high [Eu/Fe] and [Ba/Fe] together with the high
[Eu/$\alpha$] suggest that there is an extra source of r-process
elements in the Pal 1 system, beyond what is present for typical
Galactic stars.  It is unclear where the additional r-process
abundances come from; some possibilities include shot noise
(i.e. simple inhomogeneous mixing of a nearby supernova with a
progenitor in the 8-10 M$_{\sun}$ range), an unusual star formation
history that may have included an effectively truncated IMF (missing
the most massive stars that contribute significantly to the $\alpha$
elements), or even variable nucleosynthetic yields. However, each of
these possibilities has other consequences for the chemical abundance
ratios that are not clearly indicated in Pal 1.  Whatever the case,
Pal 1 has clearly not followed the ``classical'' chemical evolution
models.

Based on these chemical comparisons, we conclude that Pal 1 likely
originated in a dwarf satellite that was later accreted by the Milky
Way (about 500 Myr ago; \citealt{NiedersteOstholt}).  Though the light
elements ($Z\le 30$) are arguably in agreement with Galactic disk
stars, the neutron capture elements are distinct enough for us to
conclude that Pal 1 does have an extragalactic origin.

\subsubsection{Comparisons with Specific Galaxies: Pal 1's Host Galaxy?}\label{subsubsec:specificgalaxies}
We now attempt to chemically link Pal 1 to its host galaxy, as has
been done with e.g. Pal 12 \citep{Cohen} and Ter 7 \citep{Sbordone}.
In particular we examine the chemical abundances of the CMa
overdensity, the GASS, and the LMC intermediate-age clusters.

Given the low [La/Fe] abundance in the GASS, it is not likely that Pal
1 is a member of this stream.  The few stars analyzed in the CMa
overdensity are more promising (even if the stars analyzed so far in
CMa are more metal-rich):  the general trends in the
$\alpha$-abundances, the odd-Z elements Na, Al, Cu, and Zn, the
iron-peak elements (with the exception of Sc and V possibly due to HFS
corrections), and the neutron capture elements (possibly with the
exception of Eu) are in fair agreement.  Based on this analysis it is
possible that Pal 1 could be associated with this stream, as suggested
by \citet{Saviane} and \citet{ForbesBridges}.  The slight differences
between individual  elements could be due to the choices of spectral
lines and atomic data.  However, it must be noted that none of these
stars are guaranteed members of CMa, as disk contamination is
extremely likely \citep{SbordoneCMa}.  In particular, the star with an
[Fe/H] closest to Pal 1 seems identical to the Galactic disk stars,
and may not be a true member of the stream.  Analyses of more stars in
the CMa overdensity, particularly in the metallicity range of Pal 1,
would help establish potential membership.

A comparison with the LMC intermediate-age clusters must also be
considered, as they are also young and metal-poor.  In addition,
there are several low-mass GCs that appear to be similar to Pal 1: in
their sample of sixteen intermediate-age clusters, \citet{Milone}
found five with no signs of multiple populations in their CMDs,
suggesting that they were not massive enough to retain gas for a
second generation of stars (\citealt{ConroySpergel}; see Section
\ref{subsec:cluster}).  Though the Na, Al, Cu, and Zn abundances in
these clusters are slightly different from Pal 1, the rest of the
elements are in good agreement.  Pal 1 does not appear to have been
accreted from the LMC; however, it is possible that its formation
mechanisms were similar to the intermediate-age LMC clusters,
i.e. that Pal 1 was accreted from a dwarf galaxy similar to the LMC.

Finally, with the exceptions of [Na/Fe], [Al/Fe], [Cu/Fe], and [Zn/Fe]
(which are all higher in Pal 1), Pal 1 agrees well with Pal 12 and Ter
7 in Sgr.  Thus, regardless of its association with either the GASS or
CMa, we conclude that Pal 1 likely originated in a fairly massive
dwarf satellite (i.e. a satellite that had a mass somewhere between
Sgr and the LMC).

\section{Summary and Conclusions}\label{sec:Conclusion}
Detailed chemical abundances have been found for twenty one elements
in four red giant stars in the unusual GC Palomar 1.  Our findings are
summarized as follows:

\begin{itemize}
\item Pal 1 is an outer halo, young ($5\pm 1$ Gyr) cluster whose age,
metallicity, location, and structural parameters distinguish it from
the standard Galactic globular or open clusters.  However, Pal 1 does
appear to be similar to the known extragalactic GCs and OCs that have
been accreted during mergers (i.e. from the Sgr dSph).

\item Pal 1's young age and single stellar population suggest a
resemblance to the low-mass intermediate-age LMC clusters.  Chemical
abundances further suggest that Pal 1 may have shared a similar
formation history.

\item Pal 1 shows several unusual chemical characteristics, including
\begin{itemize}
\item Lower [$\alpha$/Fe] ratios than Galactic stars of the same
[Fe/H]
\item A lack of evidence for a Na/O anticorrelation, though marginally
high Na is found
\item Similar [Fe-peak/Fe] ratios to Galactic stars
\item An excess of second-peak s-process neutron capture elements over
first-peak elements
\item Low [Ba/Eu] and high [Eu/$\alpha$] values that suggest Pal 1's
host galaxy had an additional r-process site (possibilities could be
inhomogeneous mixing, an effectively truncated IMF, i.e. one that is
missing the most massive stars, or variable nucleosynthetic yields).
\end{itemize}

\item Chemically, Pal 1 does not behave like the typical Galactic
bulge/disk GCs or the old, metal-poor OCs.  The closest agreement
seems to be with the Sgr clusters Pal 12 and Ter 7 and the LMC
intermediate-age clusters; however the Na, Al, Cu, and Zn abundances
do not agree with the LMC or Sgr clusters.

\item Comparing the [X/Fe] of the Pal 1 stars to those of stars in
known streams show that
\begin{itemize}
\item It is unlikely that Pal 1 originated in the Galactic Anticenter
Stellar Stream, given the differing [La/Fe] ratios
\item Pal 1 may have originated in the Canis Major overdensity, if the
[X/Fe] ratios of the three possible CMa stars are extrapolated to
slightly more metal-poor stars.
\end{itemize}

\end{itemize}

Overall, we conclude that Pal 1 likely had an extragalactic origin,
though its chemistry remains unique compared to the known globular and
open clusters.

\acknowledgments
The authors wish to thank the anonymous referee for the helpful
comments and suggestions.
KAV and CS gratefully acknowledge research funding from the NSERC Discovery 
grants program.   The data presented here were collected at the Subaru
Telescope, which is operated by the National Astronomical Observatory
of Japan.
We thank M. Shetrone, V. Hill, and the DART collaboration for frequent,
helpful, and always interesting discussions on RGB analyses.   We also
thank D.A. VandenBerg, F. Herwig, and the UVic Stars Group for useful 
comments and lively discussions.  We are also grateful to D. Yong and 
P. Bonifacio for sharing their spectra and providing helpful
suggestions.
This research has made use of the NASA/ IPAC Infrared Science Archive,
which is operated by the Jet Propulsion Laboratory, California
Institute of Technology, under contract with the National Aeronautics
and Space Administration.
This research has made use of the SIMBAD database, operated at CDS,
Strasbourg, France.

{\it Facilities:}{ \facility{Subaru}}.

\end{document}